\documentclass[aps,pre,twocolumn]{revtex4-1}
\usepackage[dvips]{graphicx}
\usepackage{amsmath}
\usepackage{verbatim}
\newcommand{\etalia}{{\it et al.~}}
\newcommand{\la}{\left\langle}
\newcommand{\ra}{\right\rangle}

\newcommand{\PRL}{Phys.~Rev.~Lett.~}

\begin{document}

\title{Counterion-Induced Swelling of Ionic Microgels}

\author{Alan R. Denton}
\email[]{alan.denton@ndsu.edu}
\affiliation{Department of Physics, North Dakota State University,
Fargo, ND 58108-6050, USA}
\author{Qiyun Tang}
\altaffiliation[Current address: ]{Institut f\"ur Theoretische Physik,
Georg-August Universit\"at, 37077 G\"ottingen, Germany}
\affiliation{Department of Physics, North Dakota State University,
Fargo, ND 58108-6050, USA}

\begin{abstract}

Ionic microgel particles, when dispersed in a solvent, swell to equilibrium sizes 
that are governed by a balance between electrostatic and elastic forces.
Tuning of particle size by varying external stimuli, such as $p$H, salt concentration,
and temperature, has relevance for drug delivery, microfluidics, and filtration.  
To model swelling of ionic microgels, we derive a statistical mechanical theorem, 
which proves exact within the cell model, for the electrostatic contribution to 
the osmotic pressure inside a permeable colloidal macroion.  
Applying the theorem, we demonstrate how the distribution of counterions within 
an ionic microgel determines the internal osmotic pressure.  By combining the 
electrostatic pressure, which we compute via both Poisson-Boltzmann theory and 
molecular dynamics simulation, with the elastic pressure, modeled via the 
Flory-Rehner theory of swollen polymer networks, we show how deswelling of 
ionic microgels with increasing concentration of particles can result from a 
redistribution of counterions that reduces electrostatic pressure.  
A linearized approximation for the electrostatic pressure, which proves remarkably 
accurate, provides physical insight and greatly eases numerical calculations for 
practical applications.  Comparing with experiments, we explain why soft particles 
in deionized suspensions deswell upon increasing concentration and why this effect 
may be suppressed at higher ionic strength.
The failure of the uniform ideal-gas approximation to adequately account for 
counterion-induced deswelling below close packing of microgels is attributed 
to neglect of spatial variation of the counterion density profile and the 
electrostatic pressure of incompletely neutralized macroions.
\end{abstract}

\maketitle
\newpage


\section{Introduction}
Soft colloids, including star polymers, microgels, block-copolymer micelles, dendrimers, 
and emulsion droplets, have attracted broad attention in recent years
for their rich and tunable materials properties~\cite{vlassopoulos-cloitre2014}.
Experimental and modeling studies have explored elastic properties of 
single particles~\cite{
nieves-jcp2003,
nieves-sm2011,
cloitre-leibler1999,
cloitre-leibler2003,
nieves-macromol2000,
nieves-prl2015,
weitz-jcp2012,
tan2004,
holmqvist-shurtenberger2012,
nieves-macromol2009,
dufresne2009,
hellweg2010,
nieves-bulk-shear-pre2011,
nieves-bulk-pre2011,
nieves-sm2012,
weitz-sm2012,
ciamarra2013}
and phase behavior and dynamics of bulk suspensions~\cite{
weitz-prl1995,
groehn2000,
levin2002,
nieves-jcp2005,
winkler-gompper2012,
winkler-gompper2014,
li-chen2014,
egorov-likos2013,
colla-likos2014,
colla-likos2015,
stellbrink-likos-nanoscale2015,
stellbrink-likos-prl2015}.
Particular interest has focused on microgels~\cite{HydrogelBook2012,MicrogelBook2011,
lyon-nieves-AnnuRevPhysChem2012,yunker-yodh-review2014}
-- microscopic gel particles, composed of porous, elastic networks of cross-linked polymers,
swollen by a solvent~\cite{baker1949,pelton1986,pelton2000,saunders2009}.
Well-characterized microgels have been synthesized by 
emulsion polymerization and cross-linking of polyelectrolytes,
such as poly(N-isopropylacrylamide) (PNIPAM)~\cite{pelton2011,weitz-SM2008,yodh2013}
and poly-vinylpyridine~\cite{nieves-macromol2000,nieves-prl2015}.

When dispersed in water, microgels can acquire charge by releasing counterions into solution.
Permeability to solvent molecules and small ions drives competition between
elastic and electrostatic forces.  Swelling and equilibrium particle size
can be controlled by adjusting temperature, $p$H, and ionic strength,
leading to tunable properties
and making ionic microgels appealing for chemical sensing and 
drug delivery~\cite{hamidi2008,oh2008,oh2009,fery-AdvFunctMater2011,hoare-jcis2013}.
Recent reviews describe applications in the chemical, biomedical,
food, consumer care, pharmaceutical, and 
petroleum industries~\cite{HydrogelBook2012,MicrogelBook2011}.

Physical properties of microgel suspensions have been measured by light and small-angle 
neutron scattering, confocal microscopy, and osmometry, probing connections between 
particle elasticity, osmotic pressure, structure, and thermodynamic phase behavior~
\cite{mohanty-richtering2008,richtering2008,lyon2007,weitz-pre2012,schurtenberger-ZPC2012,
schurtenberger-SM2012,schurtenberger2013,nieves-pre2013,schurtenberger2014}.
Efforts to model ionic microgels have focused on microion distributions, 
effective electrostatic interactions between macroions, and associated thermodynamic 
phase behavior~\cite{denton2003,gottwald2005,likos2011,trizac2012,colla-likos2015,
colla-likos2014,egorov-likos2013,hedrick-chung-denton2015}.
Computer simulations of the primitive model of polyelectrolyte solutions and of 
coarse-grained bead-spring models of polyelectrolyte networks~\cite{holm2009,
molina2013,linse2002,winkler2014,winkler2016} have been essential in guiding theoretical developments.
Despite recent progress in linking single-particle properties with bulk behavior 
of ionic microgel suspensions, important challenges remain.  
In particular, the influence of counterions on the osmotic pressure inside of permeable,
compressible macroions and on equilibrium particle size is not fully understood.

While numerous experimental studies have explored the effect of salt concentration
on swelling of ionic microgels~ 
\cite{prausnitz1990,
english-grosberg1996,
nisato-langmuir1999,
nieves-jcp2001,
ortega-vinuesa2006,
nieves-salt-jpcb2008},
relatively few have examined the influence of particle concentration~ 
\cite{cloitre-leibler1999,cloitre-leibler2003,tan2004,weitz-jcp2012,
holmqvist-shurtenberger2012,nieves-prl2015}.
Borrega \etalia~\cite{cloitre-leibler1999} and Tan \etalia~\cite{tan2004} deduced
particle sizes from viscosity measurements and demonstrated that free counterions 
in solution can induce osmotic deswelling of microgels in dense suspensions.
Cloitre \etalia~\cite{cloitre-leibler2003} used dynamic light scattering (DLS) to 
quantify variations in hydrodynamic radius with cross-linker density and 
degree of ionization of polyelectrolyte microgels, synthesized from ethyl acrylate 
and methacrylic acid monomers, and proposed that free counterions may induce 
deswelling at microgel concentrations approaching close packing.
Pelaez-Fernandez \etalia~\cite{nieves-prl2015} measured the osmotic pressure 
of suspensions of cross-linked poly-vinylpyridine microgels via osmometry and dialysis. 
Over a range of concentrations, from dilute to near hard-sphere freezing, 
they obtained pressures in excess of what could be reasonably attributed to the 
microgel particles alone.  They explained their results by hypothesizing a 
dominant influence of free counterions in solution.
Like Cloitre {\it et al}., they observed significant deswelling only at 
concentrations near close packing.
Romeo \etalia~\cite{weitz-jcp2012}, using DLS to measure the hydrodynamic radius 
of relatively stiff PNIPAM particles over a range of concentrations, concluded
that any counterion-induced shrinkage is negligible.  In contrast,
Holmqvist \etalia~\cite{holmqvist-shurtenberger2012} observed much 
stronger deswelling of loosely cross-linked PNIPAM-co-PAA particles
in deionized suspensions.

Measurements of ionic microgel particle sizes are commonly interpreted in the dilute regime 
via scaling theories, originally developed for macroscopic gels~\cite{flory1953,katchalsky1951,
katchalsky1955,deGennes1979,barrat-joanny-pincus1992,rubinstein-dobrynin1996,khokhlov1997}, 
which assume strict electroneutrality and total confinement of counterions to the gel,
or via phenomenological models~\cite{cloitre-leibler1999,cloitre-leibler2003,nieves-macromol2000,
nieves-jcp2003,nieves-sm2011,nieves-prl2015,weitz-jcp2012},
which allow local deviations from electroneutrality and partial release of counterions.
Both approaches, by approximating the counterions as a uniformly distributed ideal gas, 
neglect continuous variation in the counterion density, which can be important 
for accurately modeling swelling of microscopically sized gel particles.  
To our knowledge, no previous studies have realistically modeled the direct relationship 
between the counterion density profile, osmotic pressure, and swelling of ionic microgels.

In this paper, we present a rigorous analysis of the dependence of ionic microgel size 
on counterion distribution and on the bulk concentration of particles.  By coupling 
elasticity theory of cross-linked gel networks with a new statistical mechanical theorem 
for the electrostatic contribution to the internal osmotic pressure, we demonstrate 
how the counterion distribution determines the equilibrium size of ionic microgels.  
Through a combination of theory and simulation, we explain experimentally observed 
density-dependent deswelling and identify system parameters for which such effects can be enhanced.
Thus, we predict that sufficiently soft and ionized microgels can penetrate apertures 
considerably narrower than their dilute size at concentrations below close-packing, 
with potentially important implications for drug delivery~\cite{malmsten2011}, 
microfluidics~\cite{weitz-sm2012,yunker-yodh-review2014,nordstrom2010}, and 
filtration~\cite{bacchin2014,roa-naegele2015}.

\section{Models}\label{models}
\subsection{Microgel Suspension}
We consider a suspension of soft, charged colloidal particles (macroions), permeable 
to water and microions.  Common examples are polyelectrolyte microgels and capsules 
dispersed in an aqueous electrolyte.  Within the primitive model of polyelectrolytes, 
the solvent is reduced to a dielectric continuum of dielectric constant $\epsilon$.
For simplicity, we assume $\epsilon$ to be the same inside and outside of the macroion.
This assumption can be easily relaxed to allow for nonuniform dielectric constant.
We further assume spherical macroions, of dry (collapsed) radius $a_0$, swollen radius $a$,
and charge number (valence) $Z$ associated with a fixed charge distribution, $n_f(r)$, 
which varies only with radial distance $r$ from the center.  The fixed charge comes 
from dissociation of $Z$ counterions, which can contribute to the osmotic pressure 
by virtue of their freedom to move throughout the system.  Any counterions that may be 
immobilized by condensation onto polyelectrolyte chains~\cite{manning1969} are excluded 
from this count.  It is important to note that this bare charge number can be substantially 
higher than the effective charge number associated with interparticle pair interactions, 
as is deduced from light scattering experiments~\cite{holmqvist-shurtenberger2012}.
The counterions and salt ions (microions) are modeled as point charges, of valence $\pm z$,
that can freely penetrate the macroions.  In equilibrium, the microions distribute
themselves so as to equalize the chemical potential throughout the system.  

In Donnan equilibrium, microions can exchange with a salt reservoir, of bulk microion 
pair concentration $n_0$, through an immovable, semipermeable membrane, which is 
impermeable to macroions, but permeable to both microions and solvent.
Equality of chemical potentials entails a nonuniform distribution of ions between 
the suspension and reservoir and a corresponding osmotic pressure, i.e., a difference 
in bulk pressure between the suspension and the reservoir, which is sustained by a 
counteracting force exerted by the externally fixed membrane.

Within the suspension, microions can also exchange between the interior and exterior 
macroion regions.  The periphery of a macroion thus acts analogously to a 
semipermeable membrane, allowing microions to penetrate, but holding the fixed charge 
within the macroion.  In this internal Donnan equilibrium, the fixed charge 
on the polyelectrolyte chains creates a nonuniform distribution of microions that 
equalizes the chemical potential.  The interior microion gas and the self-repulsion 
of the fixed charge within the macroion combine to generate an outward electrostatic 
pressure that swells the macroion.  Swelling is limited by the inward elastic 
restoring forces exerted by the cross-linked gel.  In equilibrium, a balance between 
these opposing pressures determines the average macroion size.  
As will become apparent below, the electrostatic contributions to the pressure 
inside and outside a microgel differ.

\begin{figure}
\begin{minipage}{4cm}
\includegraphics[width=\columnwidth,angle=0]{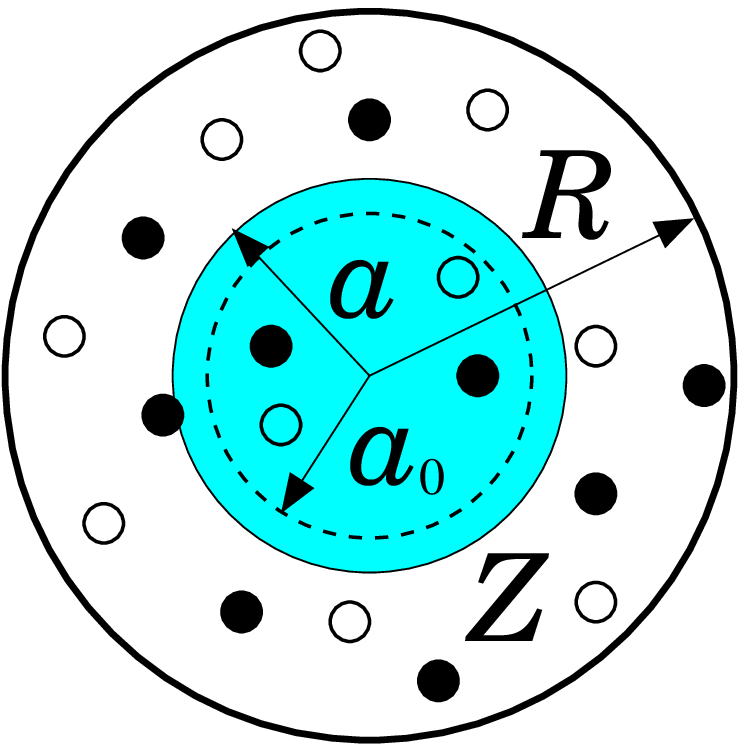}
\end{minipage}
\hspace*{0.2cm}
\begin{minipage}{4cm}
\includegraphics[width=\columnwidth,angle=0]{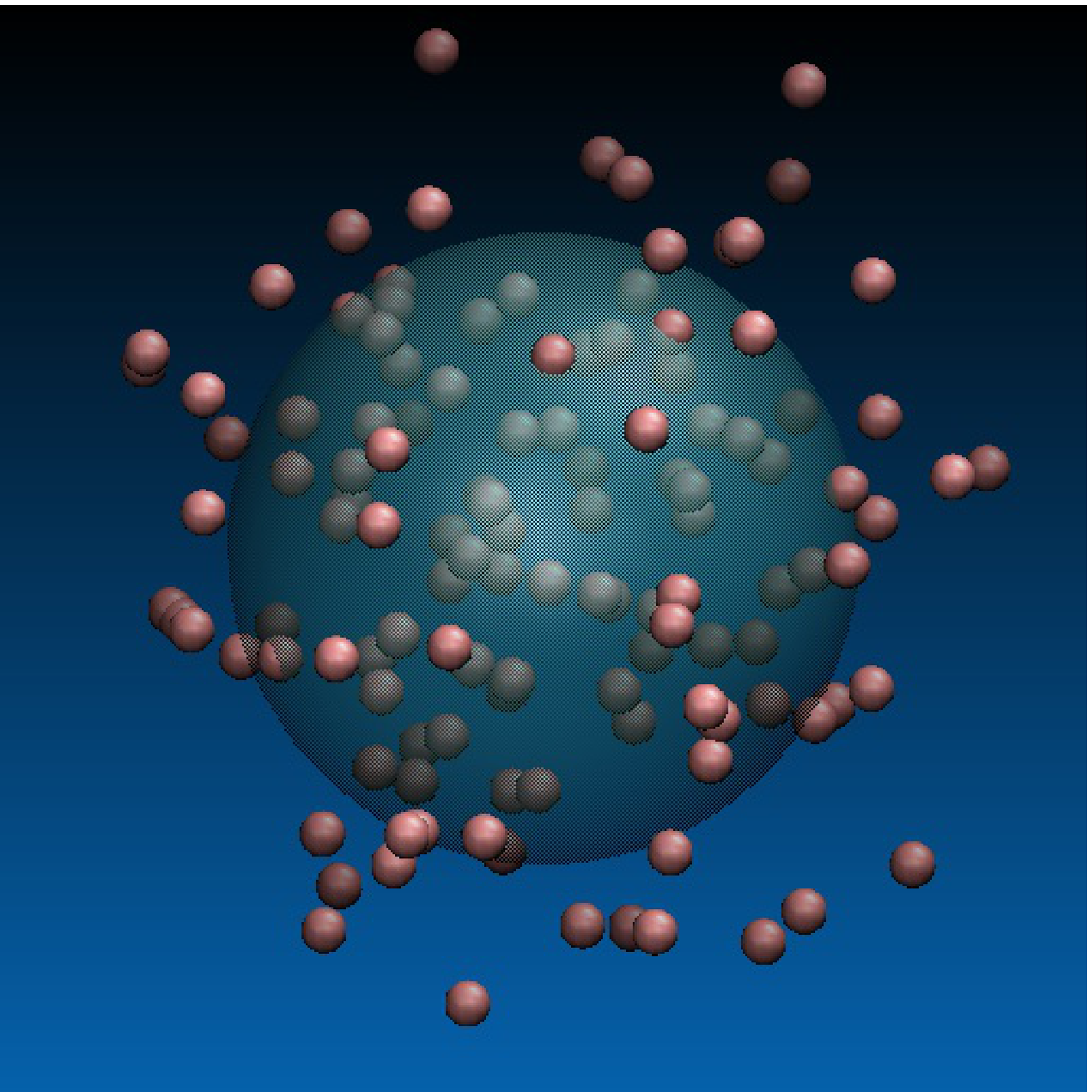}
\end{minipage}
\vspace*{-0.2cm}
\caption{
Left: Cell model with ionic microgel of swollen radius $a$ (dry radius $a_0$)
and valence $Z$ centered in a spherical cell of radius $R$ along with microions.
Right: Snapshot from MD simulation of an ionic microgel and counterions.
}\label{fig1}
\end{figure}

\subsection{Cell Model}
The cell model~\cite{marcus1955} centers a single macroion in a spherical cell 
along with $N$ mobile microions (see Fig.~\ref{fig1}).  The Hamiltonian of the system,
$H=H_e+H_g$, decouples naturally into an electrostatic component $H_e$, 
which incorporates all Coulomb interactions in the cell, and a gel component $H_g$, 
which describes the elastic and mixing degrees of freedom of a polymer network. 
The electrostatic component can be further decomposed as follows:
\begin{equation}
H_e=U_m(a)+U_{m\mu}(\{{\bf r}\};a)+U_{\mu\mu}(\{{\bf r}\})~,
\label{He}
\end{equation}
where $U_m(a)$ is the macroion self-energy and $U_{m\mu}(\{{\bf r}\};a)$ and
$U_{\mu\mu}(\{{\bf r}\})$ are the macroion-microion and microion-microion
interaction energies, respectively, which depend on the ion coordinates,
$\{{\bf r}_1,\ldots,{\bf r}_N\}\equiv\{{\bf r}\}$.  Note that only the first two terms
in Eq.~(\ref{He}) depend on the macroion radius.  The macroion-microion interaction 
may be expressed as
\begin{equation}
U_{m\mu}(\{{\bf r}\};a)=\sum_{i=1}^N v_{m\mu}({\bf r}_i;a)~,
\label{Ummu}
\end{equation}
where $v_{m\mu}({\bf r};a)$ is the macroion-microion pair potential.
Although the cell model completely neglects macroion-macroion correlations,
the relative contribution of such interparticle correlations to osmotic pressure
is known to be weakest in the low-salt limit~\cite{denton2010,hallez2014},
thus resulting in quite an accurate representation of deionized microgel 
suspensions~\cite{holm2009,molina2013,hedrick-chung-denton2015}. 
Furthermore, independent simulations of many microgel particles, interacting 
via a repulsive Hertz pair potential and fluctuating in size, reveal that for the 
same elastic parameters as considered here, macroion-macroion correlations 
and steric interactions begin to significantly affect swelling only at 
average volume fractions approaching close packing~\cite{urich-denton2016}.

\section{Theory}
A suspension that is free to exchange microions with a salt reservoir through a 
semipermeable membrane is best represented in the semi-grand canonical ensemble.
Decoupling of the electrostatic and gel components of the Hamiltonian implies a 
factorization of the semi-grand canonical partition function, $\Xi=\Xi_e{\cal Z}_g$, 
into an electrostatic grand canonical partition function $\Xi_e$ and a gel 
canonical partition function ${\cal Z}_g$.  Correspondingly, the semi-grand potential,
$\Omega=-k_BT\ln\Xi=\Omega_e+F_g$, separates conveniently into an electrostatic grand
potential, $\Omega_e=-k_BT\ln\Xi_e$, and a gel Helmholtz free energy, $F_g=-k_BT\ln{\cal Z}_g$.
We focus first on the electrostatic grand canonical partition function, in a 
spherical cell of fixed radius $R$, which can be expressed as
\begin{equation}
\Xi_e(\mu_0,a,R,T)~\propto~\sum_{N=0}^{\infty}\frac{e^{\beta\mu_0 N}}{N!}
\prod_{i=1}^N \int_0^R dr_i\, r_i^2~e^{-\beta H_e}~,
\label{Xi_e}
\end{equation}
with $\beta\equiv 1/(k_BT)$ at temperature $T$ and
$\mu_0=k_BT\ln n_0$ the microion chemical potential in the reservoir.

The bulk osmotic pressure -- the pressure in the suspension relative to the pressure 
in the reservoir -- is defined via the 
derivative of the grand potential with respect to the system volume, $V=4\pi R^3/3$:
\begin{equation}
\pi_b=-\left(\frac{\partial\Omega}{\partial V}\right)_{\mu_0,a,T}
=\frac{k_BT}{4\pi R^2}\frac{\partial}{\partial R}\ln\Xi_e(\mu_0,a,R,T)~,
\label{p01}
\end{equation}
where on the right side we have used the fact that ${\cal Z}_g$ is independent of $R$ in the cell model.
Substituting for $\Xi_e(\mu_0,a,R,T)$ from Eqs.~(\ref{He})-(\ref{Xi_e}) yields
\begin{equation}
\beta \pi_b=\la n_+(R)\ra + \la n_-(R)\ra~,
\label{p02}
\end{equation}
where $n_{\pm}(R)$ are the microion densities at the cell boundary
and $\la~\ra$ denotes an ensemble average over microion coordinates.
This classic theorem for the bulk osmotic pressure was first derived within 
PB theory~\cite{marcus1955}, but proves exact within the cell model~\cite{wennerstrom1982}.

Similarly, the internal osmotic pressure -- the pressure inside the macroions
relative to the bulk osmotic pressure -- can be defined via a derivative
of $\Omega$ with respect to the single-macroion volume, $v=4\pi a^3/3$:
\begin{equation}
\pi_{\rm in}=-\left(\frac{\partial\Omega}{\partial v}\right)_{\mu_0,R,T}
=\frac{k_BT}{4\pi a^2}\frac{\partial}{\partial a}\ln\Xi(\mu_0,a,R,T)~.
\label{p1}
\end{equation}
More explicitly, $\pi_{\rm in}$ is the excess of the osmotic pressure inside a macroion 
over the osmotic pressure $\pi_b$ outside, within the suspension.  In equilibrium, 
the electrostatic pressure is balanced by the elastic pressure of the polymer gel, 
resulting in $\pi_{\rm in}=0$.  Upon substituting $\Omega=\Omega_e+F_g$ into Eq.~(\ref{p1}), 
the internal osmotic pressure separates into electrostatic and gel contributions: 
$\pi_{\rm in}=\pi_e+\pi_g$.
From Eqs.~(\ref{He})-(\ref{Xi_e}), the electrostatic contribution 
to the internal osmotic pressure may be expressed as
\begin{equation}
\pi_e=-\frac{1}{4\pi a^2}\left(\frac{\partial}{\partial a}U_m(a)
+\la\frac{\partial}{\partial a}U_{m\mu}(a)\ra\right).
\label{p2}
\end{equation}

For the gel contribution, we invoke the Flory-Rehner theory of gel 
elasticity~\cite{flory1953,ciamarra2013}, which combines mixing entropy,
polymer-solvent interactions, and elastic network energy to predict a gel free energy
\begin{eqnarray}
\beta F_g&=&N_m\left[\left(\alpha^3-1\right)\ln\left(1-\alpha^{-3}\right)
+\chi\left(1-\alpha^{-3}\right)\right] 
\nonumber\\[1ex]
&+&\frac{3}{2}N_{\rm ch}\left(\alpha^2-\ln\alpha-1\right)~,
\label{Flory-F}
\end{eqnarray}
where $\alpha\equiv a/a_0$ is the particle swelling ratio, $N_m$ and $N_{\rm ch}$ 
are the numbers of monomers and chains per microgel, and $\chi$ is the 
Flory solvency parameter.  The corresponding gel pressure is given by
\begin{eqnarray}
\beta \pi_g v = &-&N_m\left[\alpha^3\ln\left(1-\alpha^{-3}\right) +
\chi\alpha^{-3}+1\right] \nonumber\\[1ex]
&-& N_{\rm ch}\left(\alpha^2 - 1/2\right)~.
\label{Flory-p}
\end{eqnarray}
At equilibrium swelling, the semi-grand potential is a minimum with respect to variation 
of $\alpha$, which is equivalent to vanishing of the total internal osmotic pressure: 
$\pi_{\rm in}(\alpha)=\pi_e(\alpha)+\pi_g(\alpha)=0$.  From this stability criterion, 
we explore equilibrium swelling as a function of particle concentration in Sec.~\ref{results}. 

Our theorem for the electrostatic pressure difference across the surface of a permeable 
macroion [Eq.~(\ref{p2})] is exact within the cell model.  Practical implementation now 
requires a model for the fixed charge distribution within a macroion.  To illustrate,
we proceed with the simplest model of a uniformly charged microgel of fixed charge 
number density $n_f(r)=Z/v$ ($r\le a$).  
In this case, the energy of interaction between a macroion and a microion 
of valence $z$ is given by
\begin{equation}
\beta v_{m\mu}(r)=-\frac{Zz\lambda_B}{2a}~\left(3-r^2/a^2\right), \quad r\le a~,
\label{vmmugel-sphere}
\end{equation}
and the macroion self-energy is 
\begin{equation}
\beta U_m=\frac{3}{5}Z^2\frac{\lambda_B}{a}~,
\label{umgel-sphere}
\end{equation}
where $\lambda_B\equiv e^2/(\epsilon k_BT)$ is the Bjerrum length.
From Eq.~(\ref{vmmugel-sphere}), it follows that
\begin{equation}
\beta\la\frac{\partial}{\partial a}U_{m\mu}(a)\ra
=-\frac{Z\lambda_B}{2}\la\frac{\partial}{\partial a}\sum_{i=1}^N
z_i\left(\frac{3}{a}-\frac{r_i^2}{a^3}\right)\ra~,
\label{dda}
\end{equation}
with $z_i=\pm z$ denoting the valence of microion $i$.
Now substituting Eq.~(\ref{dda}) into Eq.~(\ref{p2}) yields
\begin{equation}
\beta \pi_e v=\frac{Z\lambda_B}{2a}
\left( \frac{2}{5}Z - \la N_+\ra + \la N_-\ra
+ \frac{\la r^2\ra_+-\la r^2\ra_-}{a^2} \right).
\label{p4}
\end{equation}
where, for given radial number density profiles $n_{\pm}(r)$, 
\begin{equation}
\la N_{\pm}\ra=4\pi\int_0^a dr\, r^2 n_{\pm}(r)
\label{Npm}
\end{equation}
are the mean numbers of interior counterions/coions and
\begin{equation}
\la r^2\ra_{\pm}=4\pi \int_0^a dr\, r^4 n_{\pm}(r)
\label{r2pm}
\end{equation}
are second moments of the interior microion density profiles.
Equation~(\ref{p4}) provides an explicit formula -- exact within the 
spherical cell model -- for the electrostatic contribution to the internal 
osmotic pressure of an ionic microgel modeled as a uniformly charged sphere.  
This result may be easily generalized to other macroion architectures,
such as core-shell microgels and polyelectrolyte capsules.  
Implementing Eq.~(\ref{p4}) requires the microion density profiles inside 
of the macroion, which may be obtained from either theory or simulation.
In the next section, we discuss computational methods for computing
microion densities and osmotic pressure.

\section{Computational Methods}\label{methods}
\subsection{Nonlinear Poisson-Boltzmann Theory}
To explicitly compute the electrostatic contribution to the osmotic pressure 
internal to ionic microgel particles, we implemented Poisson-Boltzmann (PB) theory 
within the spherical cell model~\cite{marcus1955,wennerstrom1982,deserno-holm2001}.
The Poisson equation for the electrostatic potential $\psi(r)$ (in $k_BT/e$ units),
\begin{equation}
\nabla^2\psi(r)=-4\pi\lambda_B[n_+(r)-n_-(r)-n_f(r)]~,
\label{Poisson}
\end{equation}
combined with the mean-field Boltzmann approximation for the equilibrium microion densities,
\begin{equation}
n_{\pm}(r)=n_0\exp[\mp\psi(r)]~, 
\label{npm}
\end{equation}
yields the nonlinear PB equation,
\begin{equation}
\psi''(r)+\frac{2}{r}\psi'(r)=\begin{cases}
{\displaystyle \kappa_0^2\sinh\psi(r)
+\frac{3Z\lambda_B}{a^3}}~,
&0<r\le a\\[1ex]
\kappa_0^2\sinh\psi(r)~,
&a<r\le R~,
\end{cases}
\label{PBeqn}
\end{equation}
where $\kappa_0=\sqrt{8\pi\lambda_B n_0}$ is the Debye screening constant in the electrolyte
reservoir.  Solving Eq.~(\ref{PBeqn}), with boundary conditions $\psi'(0)=\psi'(R)=0$,
yields $\psi(r)$ and thus $n_{\pm}(r)$~\cite{hedrick-chung-denton2015}, from which
we compute the electrostatic pressure via Eqs.~(\ref{p4})-(\ref{r2pm}).
The electrostatic pressure may also be computed from the electrostatic grand potential,
\begin{eqnarray}
\beta\Omega_e&=&
4\pi\int_0^R dr\, r^2 \sum_{i=\pm} n_i(r)\left[\ln\left(\frac{n_i(r)}{n_0}\right)-1\right]
\nonumber\\[1ex]
&+&\frac{1}{2\lambda_B}\int_0^R dr\, r^2 |\psi'(r)|^2~,
\label{Omega_e}
\end{eqnarray}
by taking a derivative with respect to $a$:
\begin{equation}
\pi_e=-\frac{1}{4\pi a^2}\left(\frac{\partial}{\partial a}\Omega_e(\mu_0,a,R,T)\right)_{\mu_0,R,T}~.
\label{pOmega_e}
\end{equation}

\subsection{Linearized Approximation}
For comparison with the nonlinear PB theory, we also consider a linearized 
approximation that provides convenient analytical expressions.  For a suspension 
of spherical, uniformly charged microgels with average microion densities $n_{\pm}$, 
linear response theory~\cite{denton2003,hedrick-chung-denton2015} 
predicts microion density profiles (to within a constant)
\begin{equation}
n_{\pm}(r)=\pm\frac{Z}{v}\frac{n_{\pm}}{n_{\mu}}~\left\{
\begin{array} {l@{~~}l}
1-\frac{\displaystyle 1+x}{\displaystyle x}~e^{-x}
~\frac{\displaystyle a}{\displaystyle r}
\sinh\left(\frac{\displaystyle xr}{\displaystyle a}\right), &r\leq a \\[1ex]
\left(\cosh x-\frac{\displaystyle \sinh x}
{\displaystyle x}\right)\frac{\displaystyle a}{\displaystyle r}~e^{-xr/a}, &r>a~,
\end{array} \right. 
\label{nrgel}
\end{equation}
where $n_{\mu}=n_++n_-$ is the total average microion density, $x=\kappa a$, and 
$\kappa=\sqrt{4\pi\lambda_B n_{\mu}}$ is the screening constant in the suspension
({\it cf}. $\kappa_0$ in the reservoir).  For a suspension of average microgel 
and salt densities $n_m$ and $n_s$, respectively, electroneutrality dictates
$n_{\mu}=Zn_m+2n_s$.  Substituting Eq.~(\ref{nrgel}) into Eqs.~(\ref{Npm}) and 
(\ref{r2pm}) yields (to within a constant) the mean numbers of interior microions,
\begin{equation}
\la N_{\pm}\ra=\pm Z\frac{n_{\pm}}{n_{\mu}}
\left[1-3\frac{1+x}{x^3}e^{-x}\left(x\cosh x-\sinh x\right)\right]~,
\label{Npm_lin}
\end{equation}
and second moments of interior microion density profiles,
\begin{eqnarray}
&&\la r^2\ra_{\pm}=\pm\frac{3}{5}Za^2\frac{n_{\pm}}{n_{\mu}}
\bigg\{1-5\frac{1+x}{x^5}e^{-x}
\nonumber\\[1ex]
&\times&\left[x(x^2+6)\cosh x-3(x^2+2)\sinh x\right]\bigg\}~.
\label{r2pm_lin}
\end{eqnarray}
Finally, combining Eqs.~(\ref{Npm_lin}), (\ref{r2pm_lin}), and (\ref{p4}) yields
\begin{equation}
\beta \pi_e v=3Z^2\frac{\lambda_B}{a}\frac{1+x}{x^4}e^{-x}
\left(\frac{1+x^2}{x}\sinh x-3\cosh x\right)~.
\label{pe-lin}
\end{equation}
This analytical approximation, which proves to be quite accurate, greatly eases 
calculations and facilitates comparisons with experiments (see Sec.~\ref{results}).

\subsection{Molecular Dynamics Simulations}
In addition to applying PB theory, we also performed molecular dynamics (MD) 
simulations within the spherical cell model.  Using the LAMMPS molecular 
simulator~\cite{plimpton1995,lammps}, we confined a fixed number of monovalent point counterions 
-- interacting via Coulomb pair potentials -- to a spherical cell of fixed radius by 
a repulsive Lennard-Jones wall force.  We modeled the influence of the macroion on the 
counterions by imposing an ``external" electric field equal to the negative gradient 
of Eq.~(\ref{vmmugel-sphere}) and maintained a constant average temperature 
via a Nos\'e-Hoover thermostat.  The canonical (constant-$NVT$) ensemble proves more 
practical here than the grand canonical ensemble and yields the same microion distributions 
for the same system salt concentration~\cite{wennerstrom1982,holm2009,molina2013}.
Following equilibration for $10^6$ steps, we computed thermodynamic quantities by 
averaging over particle trajectories for $10^7$ time steps.  
From the resulting histogram of the counterion density, we computed $\la N_+\ra$ 
and $\la r^2\ra_+$, and then $\pi_e$ from Eq.~(\ref{p4}).

\begin{figure}[h!]
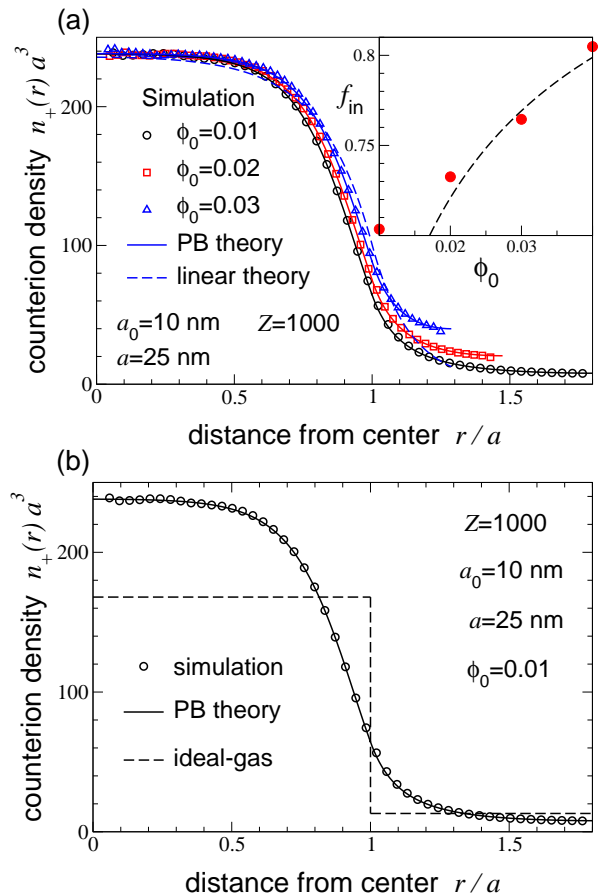

\includegraphics[width=0.9\columnwidth,angle=0]{fig2a.eps}
\includegraphics[width=0.9\columnwidth,angle=0]{fig2b.eps}
\vspace*{-0.2cm}
\caption{
(a) Counterion number density profiles near an ionic microgel of dry radius $a_0=10$ nm, 
swollen radius $a=25$ nm, and valence $Z=1000$ in a salt-free aqueous solution from 
MD simulations (symbols) and PB theory (solid curves) in the cell model 
at dry particle volume fractions $\phi_0=0.01$, 0.02, 0.03.  
Also shown is the prediction of linear response theory~\cite{denton2003} (dashed curve) 
for $\phi_0=0.03$.  Inset: Average fraction of counterions inside a macroion, 
$f_{\rm in}$ vs.~$\phi_0$, from simulations (symbols) and linear response theory (curve).
(b) Counterion density profile predicted by PB theory (solid curve) compared with 
uniform ideal-gas approximation (dashed curve).
}\label{fig2}
\end{figure}

\section{Results and Discussion}\label{results}
\subsection{Illustrative Example}
As noted above, the interior pressure theorem [Eq.~(\ref{p2})] can be applied to 
predict the electrostatic contribution to the osmotic pressure within macroions 
of any architecture, including those with nonuniform cross-linker density, such as 
core-shell~\cite{stieger2004,nieves-sm2011,weitz-jcp2012,holmqvist-shurtenberger2012,
schurtenberger-ZPC2012,ciamarra2013} or hollow~\cite{potemkin2015} microgels.
To illustrate the implementation of the theorem, we present numerical results, 
in the spherical cell model, for a uniformly charged microgel of dry radius 
$a_0=10$ nm, swollen radius $a=25$ nm, and valence $Z=1000$ in a salt-free 
aqueous solution at $T=293$ K ($\lambda_B=0.714$ nm).  
Figure~\ref{fig2}(a) shows radial profiles of monovalent counterion density 
for dry particle volume fractions $\phi_0\equiv (a_0/R)^3=0.01$, 0.02, 0.03, 
corresponding to swollen particle volume fractions 
$\phi\equiv(a/R)^3=\alpha^3\phi_0=0.156$, 0.313, 0.469.
The counterion density profiles are relatively flat near the microgel center,
where the electric field is weak, and fall off toward the periphery 
over a distance comparable to the screening length, $\kappa^{-1}$ in the
salt-free limit [see Eq.~(\ref{nrgel})].
Close agreement between simulation and theory validates the PB approximation.  
Also shown in Fig.~\ref{fig2}(a) is the linearized approximation 
for $\phi_0=0.03$, which proves quite accurate, aside from slight deviations 
near the particle periphery.
For comparison, Fig.~\ref{fig2}(b) shows the counterion density profile
assumed by the uniform ideal-gas approximation (see Sec.~\ref{results}). 

From the simulations, we also extracted the average fraction of interior mobile counterions,
$f_{\rm in}=\la N_+\ra/Z$.  Integrating the PB counterion density profile over the microgel 
volume gave nearly identical results.  The inset of Fig.~\ref{fig2} shows that 20-30\% 
of the counterions reside outside of the macroion, confirming that bulk theories of 
polyelectrolyte gels, which assume total counterion confinement, are not applicable here. 
Over this range of volume fractions ($\phi_0=$0.01-0.04, $\phi=$0.156-0.625), 
the average fraction of interior counterions is seen to increase monotonically 
and roughly linearly.  A similar trend results from the linearized approximation
[Eq.~(\ref{Npm_lin})], which proves to be accurate for $\phi_0>0.02$ ($\phi>0.3$).
Deviations at lower volume fractions stem from nonlinear screening and differences
in boundary conditions between the two theories 
(free vs.~cell boundary conditions)~\cite{denton2003,hedrick-chung-denton2015}.

Proceeding to the osmotic pressure, we computed $\pi_e$ by solving the nonlinear 
PB equation [Eq.~(\ref{PBeqn})] for $n_{\pm}(r)$ and substituting $\la N_{\pm}\ra$ 
and $\la r^2\ra_{\pm}$ into Eq.~(\ref{p4}).  
As a consistency check, we also computed $\pi_e$ numerically from 
Eq.~(\ref{pOmega_e}), using the PB solution for the counterion density profile, 
and obtained results identical to those from Eq.~(\ref{p4}).
We emphasize, however, that Eq.~(\ref{pOmega_e}), because it relies on knowledge of the 
grand potential, is in practice limited to PB theory~\cite{colla-likos2014}.
In contrast, our internal pressure theorem [Eqs.~(\ref{p2}) and (\ref{p4})]
can be used to extract the osmotic pressure also from simulations,
which naturally include correlations between microions.

\begin{figure}
\includegraphics[width=0.9\columnwidth,angle=0]{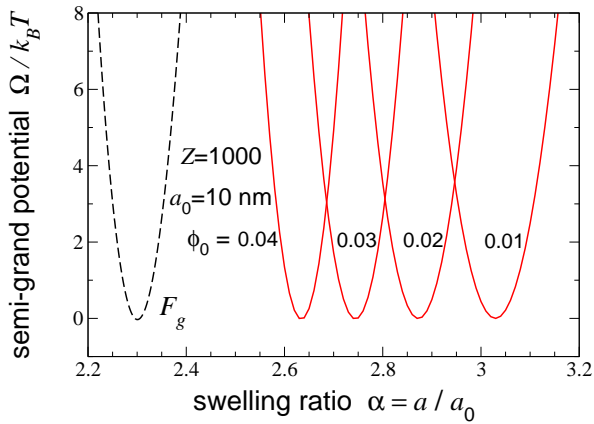}
\vspace*{-0.2cm}
\caption{
Semi-grand potential $\Omega$ vs.~particle swelling ratio $\alpha$ of ionic microgels with 
valence $Z=1000$, dry particle radius $a_0=10$ nm, $N_m=2\times 10^5$monomers, 
$N_{\rm ch}=100$ chains, and $\chi=0.5$ in deionized solutions of dry particle 
volume fractions $\phi_0=0.01-0.04$ (right to left).  To ease comparison, the minima 
are shifted to zero.  For reference, the free energy $F_g$ of a nonionic microgel 
[Eq.~(\ref{Flory-F})] is also shown (dashed curve).
}\label{fig3}
\end{figure}

To explore the equilibrium particle size, we combined the electrostatic contribution 
to the grand potential and internal osmotic pressure, which promotes microgel swelling, 
with the elastic gel contribution, which limits swelling.  
For illustration, we present results of our calculations for particles characterized 
by $a_0=10$ nm, $Z=1000$, $N_m=2\times 10^5$, $N_{\rm ch}=100$, and $\chi=0.5$, 
representing moderately charged, loosely cross-linked gels comprising monomers 
of diameter 3 \AA.  Note that the ratio $Z/N_m=0.005$ is well below the threshold 
for counterion condensation onto polyelectrolyte chains~\cite{manning1969}.  
Such relatively small microgels contain still a sufficiently high number of monomers 
to be reasonably modeled by a continuous charge distribution, especially considering 
that thermal motions of the chains tend to wash out charge discreteness.
Nevertheless, direct comparison of our predictions with data from simulations of
microscopic models with discrete charges would help to clarify any limitations 
of the continuum model.  Furthermore, our purpose here is merely to test the accuracy 
of the theory against MD simulations for the same model.  After validating the theory, 
we compare predictions with experimental data for much larger microgels carrying 
much higher charge numbers (see Sec.~\ref{comparisons}).

Variation of the Flory-Rehner free energy [Eq.~(\ref{Flory-F})] with swelling ratio 
implies thermally excited fluctuations in particle size, i.e., dynamical polydispersity.  
An isolated nonionic microgel has an equilibrium size 
that fluctuates according to a probability distribution,
\begin{equation}
P(\alpha)~\propto~\exp[-\beta F_g(\alpha)]~,
\label{FloryP}
\end{equation}
the most probable size then corresponding to the minimum of $F_g(\alpha)$.  
In the case of ionic microgels, mobile counterions generate an internal 
electrostatic pressure that modifies this size distribution and enhances swelling.

\begin{figure}
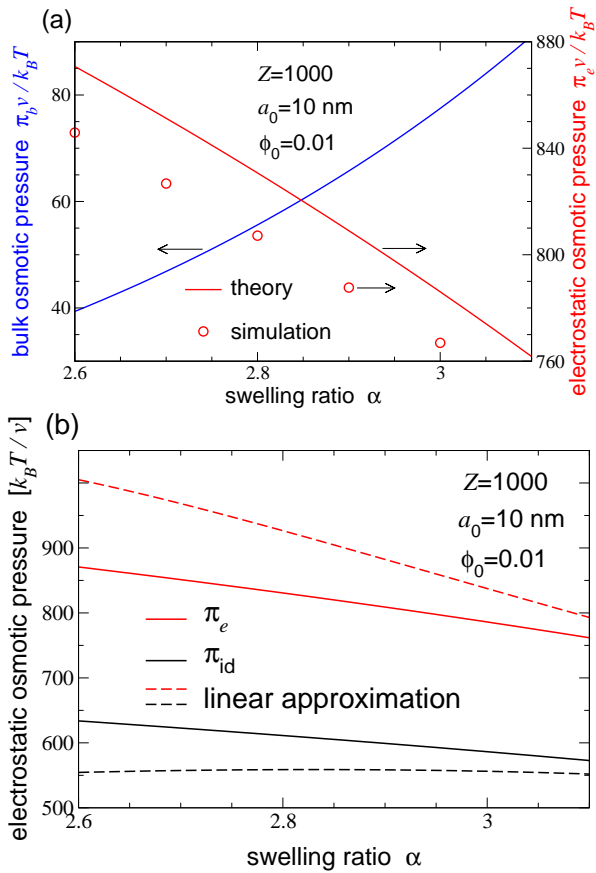

\includegraphics[width=0.9\columnwidth,angle=0]{fig4a.eps}
\includegraphics[width=0.9\columnwidth,angle=0]{fig4b.eps}
\vspace*{-0.2cm}
\caption{
(a) Osmotic pressure in bulk suspension $\pi_b$ relative to reservoir (left) [Eq.~(\ref{p02})]
and electrostatic contribution to internal osmotic pressure $\pi_e$ relative to suspension (right) 
[Eq.~(\ref{p4})] vs.~particle swelling ratio $\alpha$ for valence $Z=1000$, dry particle radius $a_0=10$ nm,
and dry particle volume fraction $\phi_0=0.01$.  Curves are from PB theory and symbols from 
MD simulations in the cell model.
(b) Electrostatic contribution to internal osmotic pressure $\pi_e$ (red curves),
uniform ideal-gas approximation $\pi_{\rm id}$ [Eq.~(\ref{pid})] (black curves), 
and linear response approximations [Eqs.~(\ref{Npm_lin}) and (\ref{pe-lin})] (dashed).
}\label{fig4}
\end{figure}

Results for the semi-grand potential [from Eqs.~(\ref{Flory-F}) and (\ref{Omega_e})]
are shown in Fig.~\ref{fig3} over a range of dry particle volume fractions.
To facilitate comparison, the minimum values of $\Omega$ are shifted to zero.
For these system parameters, the electrostatic pressure evidently produces 
significant swelling, while at the same time broadening the particle size distribution
by shifting the minimum of $\Omega(\alpha)$ out to a range of $\alpha$ with weaker curvature.
With increasing particle concentration, however, swelling is reduced and the 
size polydispersity narrows.  As the dry particle volume fraction increases 
from 0.01 to 0.04, the swollen particle volume fraction increases as well, 
despite deswelling, although more gradually, from 0.28 to 0.73.
Nevertheless, it is essential to realize that the particle deswelling and narrowing 
polydispersity predicted here are driven only by a redistribution of counterions, 
not by correlations between microgels, which are neglected in the cell model.  
At sufficiently high concentrations of microgels, electrostatic and 
steric interactions between particles will eventually affect the size distribution.

In Fig.~\ref{fig4}(a), we plot the electrostatic contribution to the internal osmotic pressure
(relative to the suspension) vs.~swelling ratio, as calculated from both theory and simulation.
For comparison, the osmotic pressure in the suspension (relative to the reservoir) is also plotted.
These results were computed via Eqs.~(\ref{p02}) and (\ref{p4}), using counterion density 
profiles calculated from PB theory and extracted as histograms from the simulations. 
For the internal electrostatic pressure, PB theory and MD simulation agree to within 3\%,
providing a consistency check and validating the mean-field Boltzmann approximation. 
With increasing swelling ratio, $\pi_b$ increases monotonically with swelling ratio, 
as the counterion density outside the microgels grows with increasing volume fraction 
(see Fig.~\ref{fig2}).  At the same time, $\pi_e$ decreases monotonically with increasing
$\alpha$, due to a declining charge density with increasing particle volume.

In Fig.~\ref{fig4}(b), we plot the total electrostatic contribution to the internal 
osmotic pressure vs.~swelling ratio together with the linear response approximation,
which proves reasonable and increasingly accurate with increasing $\alpha$ (and $\phi$).
Also shown in Fig.~\ref{fig4}(b) is the uniform ideal-gas approximation,
\begin{equation}
\beta \pi_{\rm id}v=Z\left(f_{\rm in}-(1-f_{\rm in})\frac{\phi}{1-\phi}\right)~,
\label{pid}
\end{equation}
which is commonly used as an estimate of $\pi_e$ when interpreting
experimental data~\cite{cloitre-leibler1999,nieves-macromol2000,nieves-jcp2003,
cloitre-leibler2003,nieves-sm2011,tan2004,weitz-jcp2012,holmqvist-shurtenberger2012,nieves-prl2015}.
For these parameters, this approximation is seen to significantly underestimate the 
magnitude of the electrostatic contribution to the internal osmotic pressure.  
This disparity is perhaps not surprising, given that such a relatively small microgel
is far from electroneutral.  In qualitative terms, a physical explanation for the differing 
predictions of the two approaches is that the uniform ideal-gas approximation neglects,
not only the spatial variation of the counterion density profile, as illustrated in Fig.~\ref{fig2}(b),
but also the outward electrostatic pressure of the incompletely neutralized macroion.

\begin{figure}
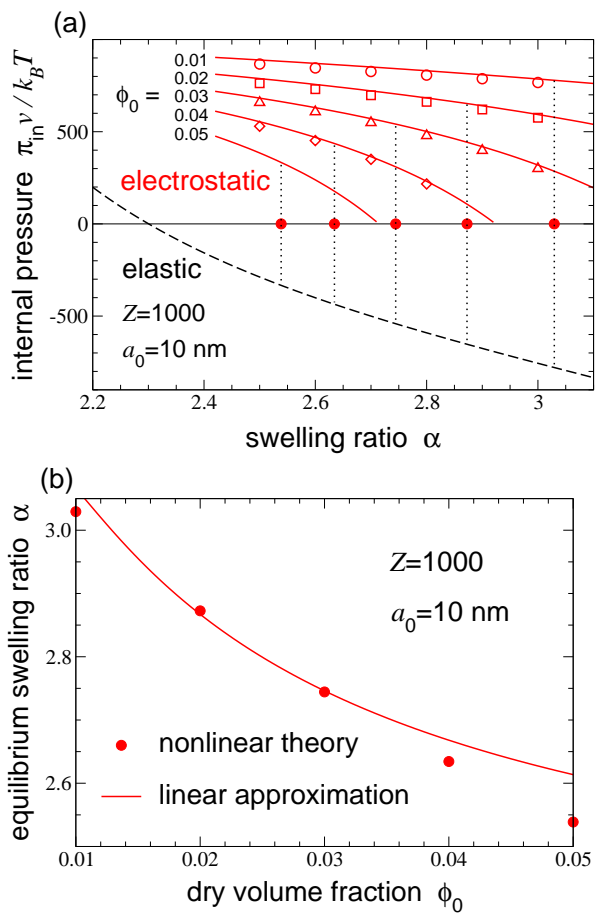

\includegraphics[width=0.9\columnwidth,angle=0]{fig5a.eps}
\\[1ex]
\includegraphics[width=0.9\columnwidth,angle=0]{fig5b.eps}
\vspace*{-0.2cm}
\caption{
(a) Contributions to pressure inside ionic microgel vs.~particle swelling ratio 
for the same system parameters as in Fig.~\ref{fig3} and dry particle volume fractions 
$\phi_0=0.01-0.05$ (top to bottom).  Electrostatic pressure $\pi_e$ [Eq.~(\ref{p4})] from 
PB theory (solid curves) and from simulations (open symbols), both in the cell model.
Elastic gel pressure $\pi_g$ (dashed curve) from Flory-Rehner theory [Eq.~(\ref{Flory-p})].
At equilibrium swelling, the total internal osmotic pressure vanishes: 
$\pi_e(\alpha)+\pi_g(\alpha)=0$ (filled symbols).
(b) Equilibrium swelling ratio vs.~$\phi_0$.  Symbols: nonlinear PB theory.
Curve: linear approximation.
}\label{fig5}
\end{figure}

The electrostatic and elastic contributions to the internal osmotic pressure are 
juxtaposed in Fig.~\ref{fig5}(a) over a range of dry volume fractions.
We computed the elastic contribution from Eq.~(\ref{Flory-p}) and the electrostatic 
contribution from Eq.~(\ref{p4}), using the counterion distributions determined from 
both PB theory [Eq.~(\ref{PBeqn})] and MD simulation.  Close agreement between theory
and simulation again provides a consistency check on our calculations.  With increasing 
dry volume fraction, the electrostatic pressure monotonically decreases, consistent with 
the shift in the semi-grand potential minimum.  This decrease in outward pressure, 
again arising from a redistribution of counterions, drives a corresponding reduction 
in equilibrium particle size.  Figure~\ref{fig5}(b) shows the equilibrium swelling ratio 
$\alpha$, computed as the root of the equation, $\pi_e(\alpha)+\pi_g(\alpha)=0$, where 
$\pi_e(\alpha)$ and $\pi_g(\alpha)$ are obtained from Eqs.~(\ref{p4})-(\ref{npm}) 
and Eq.~(\ref{Flory-p}).  For these parameters, the equilibrium swelling ratio
drops by more than 10\% from dilute to concentrated suspensions.

As a test, we also applied the linearized approximation, computing $\pi_e$ from
Eq.~(\ref{pe-lin}).  As seen in Fig.~\ref{fig5}(b), this approximation proves 
remarkably accurate. 
It is important to note that, despite the excellent agreement between PB theory and 
simulation in the present case of monovalent counterions, deviations can be expected 
for more strongly correlated multivalent counterions.  In such cases, where PB theory 
fails, the new internal osmotic pressure theorem may prove especially valuable.
Our predictions are consistent with experimental observations of weak concentration 
dependence of deswelling of relatively rigid particles~\cite{weitz-jcp2012}, but
a stronger effect for softer particles~\cite{holmqvist-shurtenberger2012}.
Next, we compare more directly with experiments.

\subsection{Comparisons with Experiments}\label{comparisons}
To illustrate the practical utility of our approach to modeling internal osmotic pressure, 
we first compare our predictions of deswelling with experimental data of Holmqvist 
\etalia~\cite{holmqvist-shurtenberger2012}, who combined static and dynamic light scattering 
with integral-equation theory and an effective pair potential model to determine the 
size and effective charge of PNIPAM-co-PAA core-shell particles as a function of 
concentration in deionized, $p$H-neutral solutions.
We choose system parameters for maximum consistency with experiments and use
corrected microgel concentrations~\cite{holmqvist-shurtenberger2012} (Erratum).
Following the prescribed limit on the number of 
dissociable groups~\cite{holmqvist-shurtenberger2012,mohanty-richtering2008}, 
we take $Z=3.5\times 10^4$.  In a deionized solution, we set the salt concentration 
to zero ($n_s=0$).  For the collapsed radius, we use the measured value of 
$a_0=50$ nm.  Consistent with particles of this size, comprised of close-packed
monomers of radius 0.3 nm, we choose the number of monomers as $N_m=3\times 10^6$.  
Again, $Z/N_m\ll 1$ precludes counterion condensation.
To best fit the shape of the distribution, we set the Flory solvency parameter 
at $\chi=0.53$, consistent with swollen polymers in water at $T=20~^{\circ}$C, and neglect 
any slight concentration dependence~\cite{nieves-sm2011}.  Lacking direct knowledge 
of the cross-linker density in the shell region, we treat $N_{\rm ch}$ as a fitting parameter.

\begin{figure}
\includegraphics[width=0.9\columnwidth,angle=0]{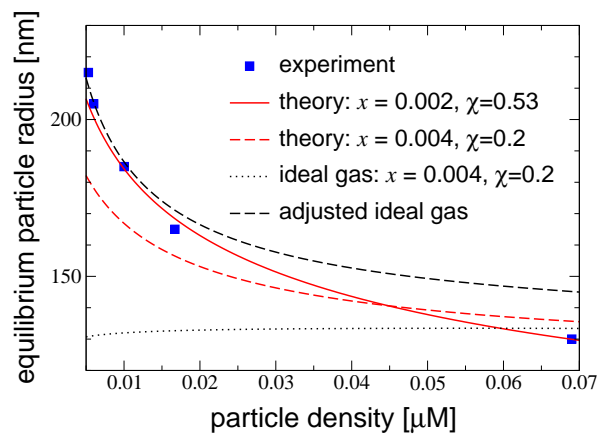}
\vspace*{-0.2cm}
\caption{
Equilibrium particle radius vs.~number density for PNIPAM-co-PAA microgels 
of valence $Z=3.5\times 10^4$, collapsed radius $a_0=50$ nm, and monomer number $N_m=3\times 10^6$
in deionized aqueous solution at $T=20~^{\circ}$C.
Theoretical predictions [Eqs.~(\ref{Flory-p}) and (\ref{pe-lin})] 
are compared with experiment~\cite{holmqvist-shurtenberger2012} (symbols)
for chain fraction
$x=N_{\rm ch}/N_m=0.002$ and solvency parameter $\chi=0.53$ (solid red curve) and for
$x=0.004$ and $\chi=0.2$ (dashed red).
Also shown, for comparison, are the
ideal-gas approximation [Eqs.~(\ref{Npm_lin}) and (\ref{pid})] (dotted black)
and a simple adjustment [Eq.~(\ref{pid-adjusted})] (dashed black)
both for $x=0.004$ and $\chi=0.2$.
From 0.005 to 0.07 $\mu$M, the volume fraction varies from $\phi\simeq 0.13$ to 0.39.
}\label{fig6}
\end{figure}

\begin{figure}
\includegraphics[width=0.9\columnwidth,angle=0]{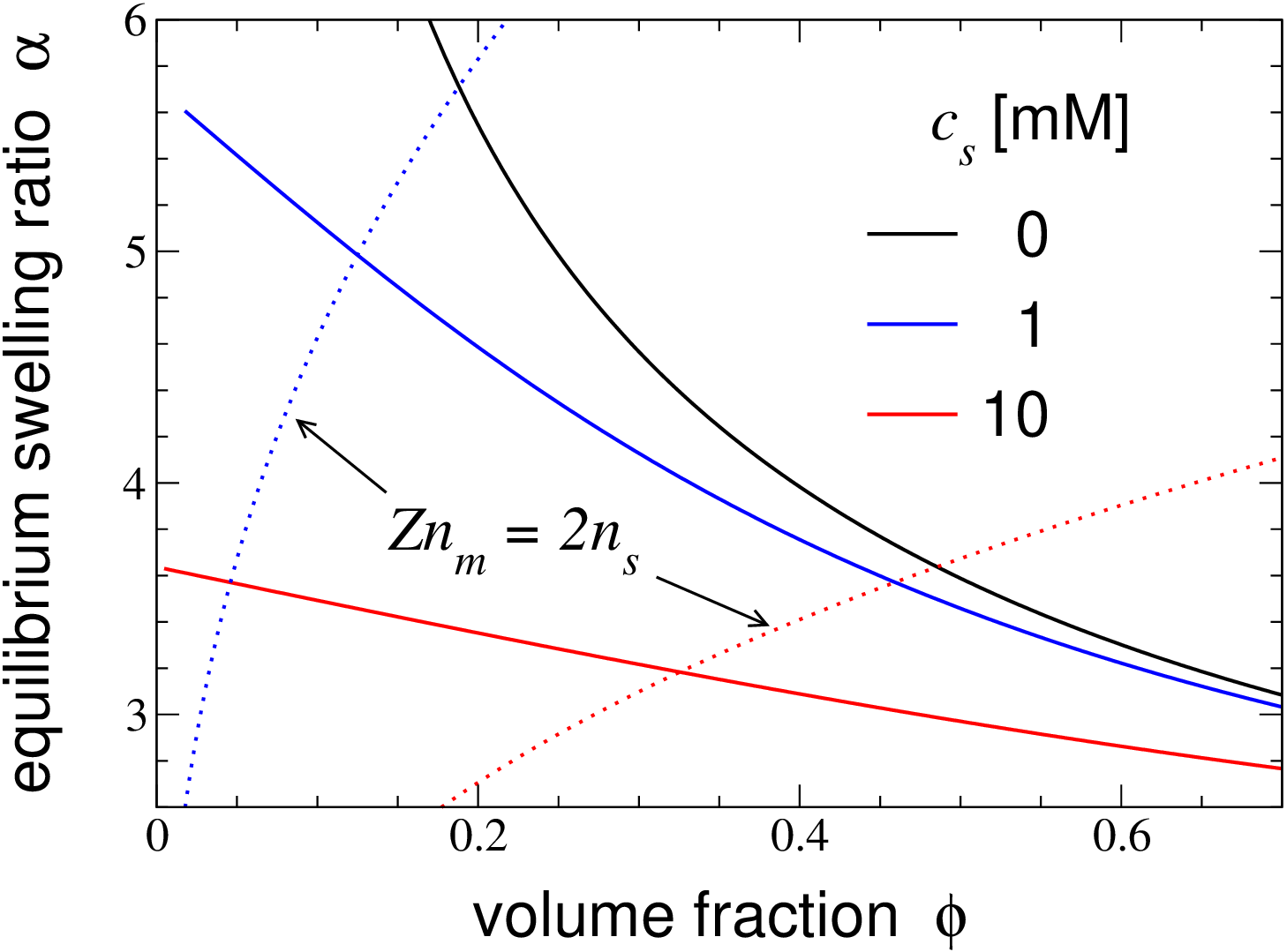}
\vspace*{-0.2cm}
\caption{
Equilibrium swelling ratio vs.~particle volume fraction for uniform-sphere model of 
poly-vinylpyridine microgels~\cite{nieves-prl2015,nieves-macromol2000} \\
of valence $Z=5\times 10^6$, 
collapsed radius $a_0=100$ nm, monomer number $N_m=2\times 10^8$, chain number $N_{\rm ch}=4\times 10^5$, 
and $\chi=0.5$ in aqueous solutions at $T=20~^{\circ}$C with system salt concentrations 
$c_s=0$ (black curve), 1 mM (blue), and 10 mM (red).  
Along dotted curves, counterion and salt ion densities are equal
for $c_s=1$ mM (blue), $c_s=10$ mM (red).
}\label{fig7}
\end{figure}

For particles of this size and charge, the nonlinear PB equation becomes so 
numerically stiff that our computational method fails to converge to a solution. 
More sophisticated iterative methods are then effective~\cite{colla-likos2014}.
However, in this parameter regime, in which most of the counterions are confined 
to the nearly electroneutral interior of the macroion, the linear response theory 
should be reasonably accurate. 
Thus, we apply the linearized approximation to model the electrostatic contribution 
to the internal osmotic pressure.  
As seen in Fig.~\ref{fig6}, we obtain a close fit to the experimental data, 
over a wide range of particle densities, with chain fraction $x\equiv N_{\rm ch}/N_m=0.002$,
which is consistent with the loosely cross-linked microgel particles in the experiments. 
To illustrate sensitivity to variation of parameters, we also show the 
prediction for $x=0.004$ and $\chi=0.2$.  
In general, the equilibrium swelling ratio decreases as $x$ and $\chi$ increase,
i.e., as the particles become stiffer and the solvent poorer.

For comparison, Fig.~\ref{fig6} also shows the prediction of the uniform ideal-gas
approximation, computed as the root of $\pi_{\rm id}+\pi_g$ with respect to $\alpha$,
using Eq.~(\ref{Npm_lin}) for the fraction of interior counterions.
Attempts to fit the data with this approximation yielded lower equilibrium radii 
and qualitatively different density dependence (dotted black curve in Fig.~\ref{fig6}),
attributable again to neglect of the macroion electrostatic pressure and to the 
relatively weak variation of $\pi_{\rm id}$ with $\alpha$ and $\phi$ [see Eq.~(\ref{pid})].
However, we find that the uniform ideal-gas approximation may be substantially 
improved by simply adding the electrostatic pressure associated with the 
self-energy of a macroion of uniformly distributed net charge
$Z_{\rm net}\equiv Z(1-f_{\rm in})$:
\begin{equation}
\beta \pi_{\rm id}v=Z\left(f_{\rm in}-(1-f_{\rm in})\frac{\phi}{1-\phi}\right)
+\frac{Z_{\rm net}^2\lambda_B}{5a}~.
\label{pid-adjusted}
\end{equation}
For comparison, predictions of Eq.~(\ref{pid-adjusted}) are also plotted 
in Fig.~\ref{fig6} (dashed black curve).
While this heuristic adjustment may prove practical for some purposes, 
our theory is more accurate and clearly more physically consistent.
Despite some potential mismatch between our model of uniformly charged macroions
and the core-shell particles studied in ref.~\cite{holmqvist-shurtenberger2012},
and some uncertainty in the variation of local $p$H inside the microgels and solvency parameter
with particle size and concentration, the level of agreement between our theory 
and experiment is encouraging and should motivate future comparisons.

While Holmqvist \etalia~\cite{holmqvist-shurtenberger2012} focused on deionized suspensions 
by working with microgels that fully ionize at neutral $p$H and flame-sealing their samples
together with an ion exchange resin, other experiments were performed at higher ionic strengths.  
For example, Borrega \etalia~\cite{cloitre-leibler1999} studied suspensions containing
substantial concentrations of sodium chloride (0.01-0.1 M), and Fern\'andez-Nieves 
\etalia~\cite{nieves-macromol2000,nieves-prl2015} studied poly-vinylpyridine microgels 
that fully ionize only at $p$H=3 (ionic strength $\sim$1 mM), achieved by adding 
sodium hydroxide.  (Note that addition of NaOH to a sample adjusts $p$H by promoting 
acid group ionization, but does not otherwise contribute to the background ion concentration.)

To explore the interrelated effects of varying both particle and salt concentrations,
we computed the equilibrium swelling ratio over a range of salt concentrations for 
parameters roughly consistent with the microgels studied in refs.~\cite{nieves-macromol2000} 
and \cite{nieves-prl2015}.  We did so by including salt concentration $n_s$ in the
Debye screening constant $\kappa$ and thus in the electrostatic contribution to the internal 
osmotic pressure $\pi_e$ [Eq.~(\ref{pe-lin})].  
At ionic strengths sufficiently high that background salt ions outnumber counterions 
dissociated from the particles ($2n_s>Zn_m$), $\kappa$ and $\pi_e$ become relatively
insensitive to changes in particle concentration.  As a consequence, with increasing 
salt concentration, not only is the degree of swelling reduced, but also the variation 
of swelling with particle density is weakened, as Fig.~\ref{fig7} illustrates.
For reference, the salt-dominated regimes ($2n_s>Zn_m$) are to the left of the 
dotted curves for respective salt concentrations.  By comparison, the uniform ideal-gas
approximation [Eq.~(\ref{pid})] predicts $\alpha\simeq 3.6$, independent of
$\phi$ and $c_s$ over these parameter ranges.  Our calculations indicate that 
high background ion concentrations may suppress counterion-induced effects and 
forestall deswelling until near close packing, where steric interactions between 
particles become significant.  Drawing conclusions about the swelling behavior
observed in refs.~\cite{nieves-macromol2000,nieves-prl2015,weitz-jcp2012} is complicated,
however, by the pronounced core-shell structure of the relatively large microgels
studied in these experiments -- 2-3 times larger than in refs.~\cite{cloitre-leibler1999}, 
\cite{cloitre-leibler2003}, \cite{tan2004}, and \cite{holmqvist-shurtenberger2012}. 
Implementing our theory for a core-shell model of microgels may help to clarify the 
origins of microgel swelling and deswelling for inhomogeneously structured microgels.

\section{Conclusions} 
Based on an exact theorem for the electrostatic contribution to the osmotic pressure 
inside a permeable macroion, we presented the first rigorous analysis of connections 
between counterion distribution, osmotic pressure, and particle swelling.
As an illustrative example, we applied the new theorem to ionic microgels,
explaining observed deswelling of particles with increasing concentration and identifying 
conditions under which deswelling and narrowing of size polydispersity can be enhanced 
via redistribution of counterions.  This electrostatically-driven phenomenon may be 
important for tuning rheological properties and facilitating microgel transport through 
narrow pores in applications ranging from drug delivery to microfluidics to filtration.  
We validated our results by comparing calculations from nonlinear Poisson-Boltzmann theory 
with data from molecular dynamics simulations in the spherical cell model.  In comparison, 
theories of macroscopic polyelectrolyte gels, which neglect both spatial variation of the 
counterion density and the electrostatic pressure of the incompletely neutralized macroion,
fail to accurately predict swelling of ionic microgels.  

For practical purposes, we also derived a linearized approximation, which provides a 
convenient analytical expression 
for the internal electrostatic pressure.  
By comparing predictions with experimental measurements of loosely cross-linked 
particles in deionized solutions, we demonstrated the ability of our theory
to explain and interpret observations of particle swelling in real microgel systems.
Our analysis demonstrates, in particular, that soft ionic microgels, when increasingly 
concentrated, can deswell due to a redistribution of counterions, and confirms that 
this unusual response can be amplified by increasing particle charge and softness 
and by minimizing ionic strength.  Moreover, we demonstrated that sensitivity of 
swelling to variations in particle density diminishes with increasing concentration 
of background ions.

Further comparisons with experiments are possible for well characterized suspensions 
of soft, ionic particles.  For consistency, however, implementation of our theory 
should be augmented to incorporate the influence of interparticle interactions 
between macroions~\cite{hedrick-chung-denton2015}, which can be important at 
concentrations approaching close-packing.  Work along these lines is in progress.

\acknowledgments
This work was supported by the National Science Foundation (Grant No.~DMR-1106331).
We thank Dr.~Jan Dhont for helpful discussions.



\begin{thebibliography}{93}%
\makeatletter
\providecommand \@ifxundefined [1]{%
 \@ifx{#1\undefined}
}%
\providecommand \@ifnum [1]{%
 \ifnum #1\expandafter \@firstoftwo
 \else \expandafter \@secondoftwo
 \fi
}%
\providecommand \@ifx [1]{%
 \ifx #1\expandafter \@firstoftwo
 \else \expandafter \@secondoftwo
 \fi
}%
\providecommand \natexlab [1]{#1}%
\providecommand \enquote  [1]{``#1''}%
\providecommand \bibnamefont  [1]{#1}%
\providecommand \bibfnamefont [1]{#1}%
\providecommand \citenamefont [1]{#1}%
\providecommand \href@noop [0]{\@secondoftwo}%
\providecommand \href [0]{\begingroup \@sanitize@url \@href}%
\providecommand \@href[1]{\@@startlink{#1}\@@href}%
\providecommand \@@href[1]{\endgroup#1\@@endlink}%
\providecommand \@sanitize@url [0]{\catcode `\\12\catcode `\$12\catcode
  `\&12\catcode `\#12\catcode `\^12\catcode `\_12\catcode `\%12\relax}%
\providecommand \@@startlink[1]{}%
\providecommand \@@endlink[0]{}%
\providecommand \url  [0]{\begingroup\@sanitize@url \@url }%
\providecommand \@url [1]{\endgroup\@href {#1}{\urlprefix }}%
\providecommand \urlprefix  [0]{URL }%
\providecommand \Eprint [0]{\href }%
\providecommand \doibase [0]{http://dx.doi.org/}%
\providecommand \selectlanguage [0]{\@gobble}%
\providecommand \bibinfo  [0]{\@secondoftwo}%
\providecommand \bibfield  [0]{\@secondoftwo}%
\providecommand \translation [1]{[#1]}%
\providecommand \BibitemOpen [0]{}%
\providecommand \bibitemStop [0]{}%
\providecommand \bibitemNoStop [0]{.\EOS\space}%
\providecommand \EOS [0]{\spacefactor3000\relax}%
\providecommand \BibitemShut  [1]{\csname bibitem#1\endcsname}%
\let\auto@bib@innerbib\@empty
\bibitem [{\citenamefont {Vlassopoulos}\ and\ \citenamefont
  {Cloitre}(2014)}]{vlassopoulos-cloitre2014}%
  \BibitemOpen
  \bibfield  {author} {\bibinfo {author} {\bibfnamefont {D.}~\bibnamefont
  {Vlassopoulos}}\ and\ \bibinfo {author} {\bibfnamefont {M.}~\bibnamefont
  {Cloitre}},\ }\href@noop {} {\bibfield  {journal} {\bibinfo  {journal} {Curr.
  Opin. Colloid Interface Sci.}\ }\textbf {\bibinfo {volume} {19}},\ \bibinfo
  {pages} {561} (\bibinfo {year} {2014})}\BibitemShut {NoStop}%
\bibitem [{\citenamefont {Fern{\'a}ndez-Nieves}\ \emph
  {et~al.}(2003)\citenamefont {Fern{\'a}ndez-Nieves}, \citenamefont
  {Fern{\'a}ndez-Barbero}, \citenamefont {Vincent},\ and\ \citenamefont {de~las
  {Nieves}}}]{nieves-jcp2003}%
  \BibitemOpen
  \bibfield  {author} {\bibinfo {author} {\bibfnamefont {A.}~\bibnamefont
  {Fern{\'a}ndez-Nieves}}, \bibinfo {author} {\bibfnamefont {A.}~\bibnamefont
  {Fern{\'a}ndez-Barbero}}, \bibinfo {author} {\bibfnamefont {B.}~\bibnamefont
  {Vincent}}, \ and\ \bibinfo {author} {\bibfnamefont {F.~J.}\ \bibnamefont
  {de~las {Nieves}}},\ }\href@noop {} {\bibfield  {journal} {\bibinfo
  {journal} {J. Chem. Phys.}\ }\textbf {\bibinfo {volume} {119}},\ \bibinfo
  {pages} {10383} (\bibinfo {year} {2003})}\BibitemShut {NoStop}%
\bibitem [{\citenamefont {Li{\'e}tor-Santos}\ \emph
  {et~al.}(2011{\natexlab{a}})\citenamefont {Li{\'e}tor-Santos}, \citenamefont
  {Sierra-Mart{\'i}n}, \citenamefont {Gasser},\ and\ \citenamefont
  {Fern{\'a}ndez-Nieves}}]{nieves-sm2011}%
  \BibitemOpen
  \bibfield  {author} {\bibinfo {author} {\bibfnamefont {J.~J.}\ \bibnamefont
  {Li{\'e}tor-Santos}}, \bibinfo {author} {\bibfnamefont {B.}~\bibnamefont
  {Sierra-Mart{\'i}n}}, \bibinfo {author} {\bibfnamefont {U.}~\bibnamefont
  {Gasser}}, \ and\ \bibinfo {author} {\bibfnamefont {A.}~\bibnamefont
  {Fern{\'a}ndez-Nieves}},\ }\href@noop {} {\bibfield  {journal} {\bibinfo
  {journal} {Soft Matter}\ }\textbf {\bibinfo {volume} {7}},\ \bibinfo {pages}
  {6370} (\bibinfo {year} {2011}{\natexlab{a}})}\BibitemShut {NoStop}%
\bibitem [{\citenamefont {Borrega}\ \emph {et~al.}(1999)\citenamefont
  {Borrega}, \citenamefont {Cloitre}, \citenamefont {Betremieux}, \citenamefont
  {Ernst},\ and\ \citenamefont {Leibler}}]{cloitre-leibler1999}%
  \BibitemOpen
  \bibfield  {author} {\bibinfo {author} {\bibfnamefont {R.}~\bibnamefont
  {Borrega}}, \bibinfo {author} {\bibfnamefont {M.}~\bibnamefont {Cloitre}},
  \bibinfo {author} {\bibfnamefont {I.}~\bibnamefont {Betremieux}}, \bibinfo
  {author} {\bibfnamefont {B.}~\bibnamefont {Ernst}}, \ and\ \bibinfo {author}
  {\bibfnamefont {L.}~\bibnamefont {Leibler}},\ }\href@noop {} {\bibfield
  {journal} {\bibinfo  {journal} {Euro. Phys. Lett.}\ }\textbf {\bibinfo
  {volume} {47}},\ \bibinfo {pages} {729} (\bibinfo {year} {1999})}\BibitemShut
  {NoStop}%
\bibitem [{\citenamefont {Cloitre}\ \emph {et~al.}(2003)\citenamefont
  {Cloitre}, \citenamefont {Borrega}, \citenamefont {Monti},\ and\
  \citenamefont {Leibler}}]{cloitre-leibler2003}%
  \BibitemOpen
  \bibfield  {author} {\bibinfo {author} {\bibfnamefont {M.}~\bibnamefont
  {Cloitre}}, \bibinfo {author} {\bibfnamefont {R.}~\bibnamefont {Borrega}},
  \bibinfo {author} {\bibfnamefont {F.}~\bibnamefont {Monti}}, \ and\ \bibinfo
  {author} {\bibfnamefont {L.}~\bibnamefont {Leibler}},\ }\href@noop {}
  {\bibfield  {journal} {\bibinfo  {journal} {C. R. Physique}\ }\textbf
  {\bibinfo {volume} {4}},\ \bibinfo {pages} {221} (\bibinfo {year}
  {2003})}\BibitemShut {NoStop}%
\bibitem [{\citenamefont {Fern{\'a}ndez-Nieves}\ \emph
  {et~al.}(2000)\citenamefont {Fern{\'a}ndez-Nieves}, \citenamefont
  {Fern{\'a}ndez-Barbero}, \citenamefont {Vincent},\ and\ \citenamefont {de~las
  {Nieves}}}]{nieves-macromol2000}%
  \BibitemOpen
  \bibfield  {author} {\bibinfo {author} {\bibfnamefont {A.}~\bibnamefont
  {Fern{\'a}ndez-Nieves}}, \bibinfo {author} {\bibfnamefont {A.}~\bibnamefont
  {Fern{\'a}ndez-Barbero}}, \bibinfo {author} {\bibfnamefont {B.}~\bibnamefont
  {Vincent}}, \ and\ \bibinfo {author} {\bibfnamefont {F.~J.}\ \bibnamefont
  {de~las {Nieves}}},\ }\href@noop {} {\bibfield  {journal} {\bibinfo
  {journal} {Macromol.}\ }\textbf {\bibinfo {volume} {33}},\ \bibinfo {pages}
  {2114} (\bibinfo {year} {2000})}\BibitemShut {NoStop}%
\bibitem [{\citenamefont {Pelaez-Fernandez}\ \emph {et~al.}(2015)\citenamefont
  {Pelaez-Fernandez}, \citenamefont {Souslov}, \citenamefont {Lyon},
  \citenamefont {Goldbart},\ and\ \citenamefont
  {Fern{\'a}ndez-Nieves}}]{nieves-prl2015}%
  \BibitemOpen
  \bibfield  {author} {\bibinfo {author} {\bibfnamefont {M.}~\bibnamefont
  {Pelaez-Fernandez}}, \bibinfo {author} {\bibfnamefont {A.}~\bibnamefont
  {Souslov}}, \bibinfo {author} {\bibfnamefont {L.~A.}\ \bibnamefont {Lyon}},
  \bibinfo {author} {\bibfnamefont {P.~M.}\ \bibnamefont {Goldbart}}, \ and\
  \bibinfo {author} {\bibfnamefont {A.}~\bibnamefont {Fern{\'a}ndez-Nieves}},\
  }\href@noop {} {\bibfield  {journal} {\bibinfo  {journal} {\PRL}\ }\textbf
  {\bibinfo {volume} {114}},\ \bibinfo {pages} {098303} (\bibinfo {year}
  {2015})}\BibitemShut {NoStop}%
\bibitem [{\citenamefont {Romeo}\ \emph {et~al.}(2012)\citenamefont {Romeo},
  \citenamefont {Imperiali}, \citenamefont {Kim}, \citenamefont
  {Fern{\'a}ndez-Nieves},\ and\ \citenamefont {Weitz}}]{weitz-jcp2012}%
  \BibitemOpen
  \bibfield  {author} {\bibinfo {author} {\bibfnamefont {G.}~\bibnamefont
  {Romeo}}, \bibinfo {author} {\bibfnamefont {L.}~\bibnamefont {Imperiali}},
  \bibinfo {author} {\bibfnamefont {J.-W.}\ \bibnamefont {Kim}}, \bibinfo
  {author} {\bibfnamefont {A.}~\bibnamefont {Fern{\'a}ndez-Nieves}}, \ and\
  \bibinfo {author} {\bibfnamefont {D.~A.}\ \bibnamefont {Weitz}},\ }\href@noop
  {} {\bibfield  {journal} {\bibinfo  {journal} {J. Chem. Phys.}\ }\textbf
  {\bibinfo {volume} {136}},\ \bibinfo {pages} {124905} (\bibinfo {year}
  {2012})}\BibitemShut {NoStop}%
\bibitem [{\citenamefont {Tan}\ \emph {et~al.}(2004)\citenamefont {Tan},
  \citenamefont {Tam}, \citenamefont {Lam},\ and\ \citenamefont
  {Tan}}]{tan2004}%
  \BibitemOpen
  \bibfield  {author} {\bibinfo {author} {\bibfnamefont {B.~H.}\ \bibnamefont
  {Tan}}, \bibinfo {author} {\bibfnamefont {K.~C.}\ \bibnamefont {Tam}},
  \bibinfo {author} {\bibfnamefont {Y.~C.}\ \bibnamefont {Lam}}, \ and\
  \bibinfo {author} {\bibfnamefont {C.~B.}\ \bibnamefont {Tan}},\ }\href@noop
  {} {\bibfield  {journal} {\bibinfo  {journal} {J. Rheol.}\ }\textbf {\bibinfo
  {volume} {48}},\ \bibinfo {pages} {915} (\bibinfo {year} {2004})}\BibitemShut
  {NoStop}%
\bibitem [{\citenamefont {Holmqvist}\ \emph {et~al.}(2012)\citenamefont
  {Holmqvist}, \citenamefont {Mohanty}, \citenamefont {N{\"a}gele},
  \citenamefont {Schurtenberger},\ and\ \citenamefont
  {Heinen}}]{holmqvist-shurtenberger2012}%
  \BibitemOpen
  \bibfield  {author} {\bibinfo {author} {\bibfnamefont {P.}~\bibnamefont
  {Holmqvist}}, \bibinfo {author} {\bibfnamefont {P.~S.}\ \bibnamefont
  {Mohanty}}, \bibinfo {author} {\bibfnamefont {G.}~\bibnamefont {N{\"a}gele}},
  \bibinfo {author} {\bibfnamefont {P.}~\bibnamefont {Schurtenberger}}, \ and\
  \bibinfo {author} {\bibfnamefont {M.}~\bibnamefont {Heinen}},\ }\href@noop {}
  {\bibfield  {journal} {\bibinfo  {journal} {\PRL}\ }\textbf {\bibinfo
  {volume} {109}},\ \bibinfo {pages} {048302} (\bibinfo {year} {2012})},\
  \bibinfo {note} {{E}rratum, in press (2016)}\BibitemShut {NoStop}%
\bibitem [{\citenamefont {Li{\'e}tor-Santos}\ \emph {et~al.}(2009)\citenamefont
  {Li{\'e}tor-Santos}, \citenamefont {Sierra-Mart{\'i}n}, \citenamefont
  {Vavrin}, \citenamefont {Hu}, \citenamefont {Gasser},\ and\ \citenamefont
  {Fern{\'a}ndez-Nieves}}]{nieves-macromol2009}%
  \BibitemOpen
  \bibfield  {author} {\bibinfo {author} {\bibfnamefont {J.~J.}\ \bibnamefont
  {Li{\'e}tor-Santos}}, \bibinfo {author} {\bibfnamefont {B.}~\bibnamefont
  {Sierra-Mart{\'i}n}}, \bibinfo {author} {\bibfnamefont {R.}~\bibnamefont
  {Vavrin}}, \bibinfo {author} {\bibfnamefont {Z.}~\bibnamefont {Hu}}, \bibinfo
  {author} {\bibfnamefont {U.}~\bibnamefont {Gasser}}, \ and\ \bibinfo {author}
  {\bibfnamefont {A.}~\bibnamefont {Fern{\'a}ndez-Nieves}},\ }\href@noop {}
  {\bibfield  {journal} {\bibinfo  {journal} {Macromol.}\ }\textbf {\bibinfo
  {volume} {42}},\ \bibinfo {pages} {6225} (\bibinfo {year}
  {2009})}\BibitemShut {NoStop}%
\bibitem [{\citenamefont {Hashmi}\ and\ \citenamefont
  {Dufresne}(2009)}]{dufresne2009}%
  \BibitemOpen
  \bibfield  {author} {\bibinfo {author} {\bibfnamefont {S.~M.}\ \bibnamefont
  {Hashmi}}\ and\ \bibinfo {author} {\bibfnamefont {E.~R.}\ \bibnamefont
  {Dufresne}},\ }\href@noop {} {\bibfield  {journal} {\bibinfo  {journal} {Soft
  Matter}\ }\textbf {\bibinfo {volume} {5}},\ \bibinfo {pages} {3682} (\bibinfo
  {year} {2009})}\BibitemShut {NoStop}%
\bibitem [{\citenamefont {Hertle}\ \emph {et~al.}(2010)\citenamefont {Hertle},
  \citenamefont {Zeiser}, \citenamefont {Hasen{\"o}hrl}, \citenamefont
  {Busch},\ and\ \citenamefont {Hellweg}}]{hellweg2010}%
  \BibitemOpen
  \bibfield  {author} {\bibinfo {author} {\bibfnamefont {Y.}~\bibnamefont
  {Hertle}}, \bibinfo {author} {\bibfnamefont {M.}~\bibnamefont {Zeiser}},
  \bibinfo {author} {\bibfnamefont {C.}~\bibnamefont {Hasen{\"o}hrl}}, \bibinfo
  {author} {\bibfnamefont {P.}~\bibnamefont {Busch}}, \ and\ \bibinfo {author}
  {\bibfnamefont {T.}~\bibnamefont {Hellweg}},\ }\href@noop {} {\bibfield
  {journal} {\bibinfo  {journal} {Colloid Polym. Sci.}\ }\textbf {\bibinfo
  {volume} {288}},\ \bibinfo {pages} {1047} (\bibinfo {year}
  {2010})}\BibitemShut {NoStop}%
\bibitem [{\citenamefont {Li{\'e}tor-Santos}\ \emph
  {et~al.}(2011{\natexlab{b}})\citenamefont {Li{\'e}tor-Santos}, \citenamefont
  {Sierra-Mart{\'i}n},\ and\ \citenamefont
  {Fern{\'a}ndez-Nieves}}]{nieves-bulk-shear-pre2011}%
  \BibitemOpen
  \bibfield  {author} {\bibinfo {author} {\bibfnamefont {J.~J.}\ \bibnamefont
  {Li{\'e}tor-Santos}}, \bibinfo {author} {\bibfnamefont {B.}~\bibnamefont
  {Sierra-Mart{\'i}n}}, \ and\ \bibinfo {author} {\bibfnamefont
  {A.}~\bibnamefont {Fern{\'a}ndez-Nieves}},\ }\href@noop {} {\bibfield
  {journal} {\bibinfo  {journal} {Phys. Rev. E}\ }\textbf {\bibinfo {volume}
  {84}},\ \bibinfo {pages} {060402(R)} (\bibinfo {year}
  {2011}{\natexlab{b}})}\BibitemShut {NoStop}%
\bibitem [{\citenamefont {Sierra-Mart{\'i}n}\ \emph {et~al.}(2011)\citenamefont
  {Sierra-Mart{\'i}n}, \citenamefont {Laporte}, \citenamefont {South},
  \citenamefont {Lyon},\ and\ \citenamefont
  {Fern{\'a}ndez-Nieves}}]{nieves-bulk-pre2011}%
  \BibitemOpen
  \bibfield  {author} {\bibinfo {author} {\bibfnamefont {B.}~\bibnamefont
  {Sierra-Mart{\'i}n}}, \bibinfo {author} {\bibfnamefont {Y.}~\bibnamefont
  {Laporte}}, \bibinfo {author} {\bibfnamefont {A.~B.}\ \bibnamefont {South}},
  \bibinfo {author} {\bibfnamefont {L.~A.}\ \bibnamefont {Lyon}}, \ and\
  \bibinfo {author} {\bibfnamefont {A.}~\bibnamefont {Fern{\'a}ndez-Nieves}},\
  }\href@noop {} {\bibfield  {journal} {\bibinfo  {journal} {Phys. Rev. E}\
  }\textbf {\bibinfo {volume} {84}},\ \bibinfo {pages} {011406} (\bibinfo
  {year} {2011})}\BibitemShut {NoStop}%
\bibitem [{\citenamefont {Sierra-Mart{\'i}n}\ and\ \citenamefont
  {Fern{\'a}ndez-Nieves}(2012)}]{nieves-sm2012}%
  \BibitemOpen
  \bibfield  {author} {\bibinfo {author} {\bibfnamefont {B.}~\bibnamefont
  {Sierra-Mart{\'i}n}}\ and\ \bibinfo {author} {\bibfnamefont {A.}~\bibnamefont
  {Fern{\'a}ndez-Nieves}},\ }\href@noop {} {\bibfield  {journal} {\bibinfo
  {journal} {Soft Matter}\ }\textbf {\bibinfo {volume} {8}},\ \bibinfo {pages}
  {4141} (\bibinfo {year} {2012})}\BibitemShut {NoStop}%
\bibitem [{\citenamefont {Menut}\ \emph {et~al.}(2012)\citenamefont {Menut},
  \citenamefont {Seiffert}, \citenamefont {Sprakel},\ and\ \citenamefont
  {Weitz}}]{weitz-sm2012}%
  \BibitemOpen
  \bibfield  {author} {\bibinfo {author} {\bibfnamefont {P.}~\bibnamefont
  {Menut}}, \bibinfo {author} {\bibfnamefont {S.}~\bibnamefont {Seiffert}},
  \bibinfo {author} {\bibfnamefont {J.}~\bibnamefont {Sprakel}}, \ and\
  \bibinfo {author} {\bibfnamefont {D.~A.}\ \bibnamefont {Weitz}},\ }\href@noop
  {} {\bibfield  {journal} {\bibinfo  {journal} {Soft Matter}\ }\textbf
  {\bibinfo {volume} {8}},\ \bibinfo {pages} {156} (\bibinfo {year}
  {2012})}\BibitemShut {NoStop}%
\bibitem [{\citenamefont {Romeo}\ and\ \citenamefont
  {Ciamarra}(2013)}]{ciamarra2013}%
  \BibitemOpen
  \bibfield  {author} {\bibinfo {author} {\bibfnamefont {G.}~\bibnamefont
  {Romeo}}\ and\ \bibinfo {author} {\bibfnamefont {M.~P.}\ \bibnamefont
  {Ciamarra}},\ }\href@noop {} {\bibfield  {journal} {\bibinfo  {journal} {Soft
  Matter}\ }\textbf {\bibinfo {volume} {9}},\ \bibinfo {pages} {5401} (\bibinfo
  {year} {2013})}\BibitemShut {NoStop}%
\bibitem [{\citenamefont {Mason}\ \emph {et~al.}(1995)\citenamefont {Mason},
  \citenamefont {Bibette},\ and\ \citenamefont {Weitz}}]{weitz-prl1995}%
  \BibitemOpen
  \bibfield  {author} {\bibinfo {author} {\bibfnamefont {T.~G.}\ \bibnamefont
  {Mason}}, \bibinfo {author} {\bibfnamefont {J.}~\bibnamefont {Bibette}}, \
  and\ \bibinfo {author} {\bibfnamefont {D.~A.}\ \bibnamefont {Weitz}},\
  }\href@noop {} {\bibfield  {journal} {\bibinfo  {journal} {\PRL}\ }\textbf
  {\bibinfo {volume} {75}},\ \bibinfo {pages} {2051} (\bibinfo {year}
  {1995})}\BibitemShut {NoStop}%
\bibitem [{\citenamefont {Gr{\"o}hn}\ and\ \citenamefont
  {Antonietti}(2000)}]{groehn2000}%
  \BibitemOpen
  \bibfield  {author} {\bibinfo {author} {\bibfnamefont {F.}~\bibnamefont
  {Gr{\"o}hn}}\ and\ \bibinfo {author} {\bibfnamefont {M.}~\bibnamefont
  {Antonietti}},\ }\href@noop {} {\bibfield  {journal} {\bibinfo  {journal}
  {Macromol.}\ }\textbf {\bibinfo {volume} {33}},\ \bibinfo {pages} {5938}
  (\bibinfo {year} {2000})}\BibitemShut {NoStop}%
\bibitem [{\citenamefont {Levin}(2002)}]{levin2002}%
  \BibitemOpen
  \bibfield  {author} {\bibinfo {author} {\bibfnamefont {Y.}~\bibnamefont
  {Levin}},\ }\href@noop {} {\bibfield  {journal} {\bibinfo  {journal} {Phys.
  Rev. E}\ }\textbf {\bibinfo {volume} {65}},\ \bibinfo {pages} {036143}
  (\bibinfo {year} {2002})}\BibitemShut {NoStop}%
\bibitem [{\citenamefont {Fern{\'a}ndez-Nieves}\ and\ \citenamefont
  {M{\'a}rquez}(2005)}]{nieves-jcp2005}%
  \BibitemOpen
  \bibfield  {author} {\bibinfo {author} {\bibfnamefont {A.}~\bibnamefont
  {Fern{\'a}ndez-Nieves}}\ and\ \bibinfo {author} {\bibfnamefont
  {M.}~\bibnamefont {M{\'a}rquez}},\ }\href@noop {} {\bibfield  {journal}
  {\bibinfo  {journal} {J. Chem. Phys.}\ }\textbf {\bibinfo {volume} {122}},\
  \bibinfo {pages} {084702} (\bibinfo {year} {2005})}\BibitemShut {NoStop}%
\bibitem [{\citenamefont {Singh}\ \emph {et~al.}(2012)\citenamefont {Singh},
  \citenamefont {Fedosov}, \citenamefont {Chatterji}, \citenamefont {Winkler},\
  and\ \citenamefont {Gompper}}]{winkler-gompper2012}%
  \BibitemOpen
  \bibfield  {author} {\bibinfo {author} {\bibfnamefont {S.~P.}\ \bibnamefont
  {Singh}}, \bibinfo {author} {\bibfnamefont {D.~A.}\ \bibnamefont {Fedosov}},
  \bibinfo {author} {\bibfnamefont {A.}~\bibnamefont {Chatterji}}, \bibinfo
  {author} {\bibfnamefont {R.~G.}\ \bibnamefont {Winkler}}, \ and\ \bibinfo
  {author} {\bibfnamefont {G.}~\bibnamefont {Gompper}},\ }\href@noop {}
  {\bibfield  {journal} {\bibinfo  {journal} {J. Phys.: Condens. Matter}\
  }\textbf {\bibinfo {volume} {24}},\ \bibinfo {pages} {464103} (\bibinfo
  {year} {2012})}\BibitemShut {NoStop}%
\bibitem [{\citenamefont {Winkler}\ \emph {et~al.}(2014)\citenamefont
  {Winkler}, \citenamefont {Fedosov},\ and\ \citenamefont
  {Gompper}}]{winkler-gompper2014}%
  \BibitemOpen
  \bibfield  {author} {\bibinfo {author} {\bibfnamefont {R.~G.}\ \bibnamefont
  {Winkler}}, \bibinfo {author} {\bibfnamefont {D.~A.}\ \bibnamefont
  {Fedosov}}, \ and\ \bibinfo {author} {\bibfnamefont {G.}~\bibnamefont
  {Gompper}},\ }\href@noop {} {\bibfield  {journal} {\bibinfo  {journal} {Curr.
  Opin. Colloid Interface Sci.}\ }\textbf {\bibinfo {volume} {19}},\ \bibinfo
  {pages} {594} (\bibinfo {year} {2014})}\BibitemShut {NoStop}%
\bibitem [{\citenamefont {Li}\ \emph {et~al.}(2014)\citenamefont {Li},
  \citenamefont {S{\'a}nchez-Di{\'a}z}, \citenamefont {Wu}, \citenamefont
  {Hamilton}, \citenamefont {Falus}, \citenamefont {Porcar}, \citenamefont
  {Liu}, \citenamefont {Do}, \citenamefont {Faraone}, \citenamefont {Smith},
  \citenamefont {Egami},\ and\ \citenamefont {Chen}}]{li-chen2014}%
  \BibitemOpen
  \bibfield  {author} {\bibinfo {author} {\bibfnamefont {X.}~\bibnamefont
  {Li}}, \bibinfo {author} {\bibfnamefont {L.~E.}\ \bibnamefont
  {S{\'a}nchez-Di{\'a}z}}, \bibinfo {author} {\bibfnamefont {B.}~\bibnamefont
  {Wu}}, \bibinfo {author} {\bibfnamefont {W.~A.}\ \bibnamefont {Hamilton}},
  \bibinfo {author} {\bibfnamefont {P.}~\bibnamefont {Falus}}, \bibinfo
  {author} {\bibfnamefont {L.}~\bibnamefont {Porcar}}, \bibinfo {author}
  {\bibfnamefont {Y.}~\bibnamefont {Liu}}, \bibinfo {author} {\bibfnamefont
  {C.}~\bibnamefont {Do}}, \bibinfo {author} {\bibfnamefont {A.}~\bibnamefont
  {Faraone}}, \bibinfo {author} {\bibfnamefont {G.~S.}\ \bibnamefont {Smith}},
  \bibinfo {author} {\bibfnamefont {T.}~\bibnamefont {Egami}}, \ and\ \bibinfo
  {author} {\bibfnamefont {W.-R.}\ \bibnamefont {Chen}},\ }\href@noop {}
  {\bibfield  {journal} {\bibinfo  {journal} {ACS Macro Lett.}\ }\textbf
  {\bibinfo {volume} {3}},\ \bibinfo {pages} {1271} (\bibinfo {year}
  {2014})}\BibitemShut {NoStop}%
\bibitem [{\citenamefont {Egorov}\ \emph {et~al.}(2013)\citenamefont {Egorov},
  \citenamefont {Paturej}, \citenamefont {Likos},\ and\ \citenamefont
  {Milchev}}]{egorov-likos2013}%
  \BibitemOpen
  \bibfield  {author} {\bibinfo {author} {\bibfnamefont {S.~A.}\ \bibnamefont
  {Egorov}}, \bibinfo {author} {\bibfnamefont {J.}~\bibnamefont {Paturej}},
  \bibinfo {author} {\bibfnamefont {C.~N.}\ \bibnamefont {Likos}}, \ and\
  \bibinfo {author} {\bibfnamefont {A.}~\bibnamefont {Milchev}},\ }\href@noop
  {} {\bibfield  {journal} {\bibinfo  {journal} {Macromol.}\ }\textbf {\bibinfo
  {volume} {46}},\ \bibinfo {pages} {3648} (\bibinfo {year}
  {2013})}\BibitemShut {NoStop}%
\bibitem [{\citenamefont {Colla}\ \emph {et~al.}(2014)\citenamefont {Colla},
  \citenamefont {Likos},\ and\ \citenamefont {Levin}}]{colla-likos2014}%
  \BibitemOpen
  \bibfield  {author} {\bibinfo {author} {\bibfnamefont {T.}~\bibnamefont
  {Colla}}, \bibinfo {author} {\bibfnamefont {C.~N.}\ \bibnamefont {Likos}}, \
  and\ \bibinfo {author} {\bibfnamefont {Y.}~\bibnamefont {Levin}},\
  }\href@noop {} {\bibfield  {journal} {\bibinfo  {journal} {J. Chem. Phys.}\
  }\textbf {\bibinfo {volume} {141}},\ \bibinfo {pages} {234902} (\bibinfo
  {year} {2014})}\BibitemShut {NoStop}%
\bibitem [{\citenamefont {Colla}\ and\ \citenamefont
  {Likos}(2015)}]{colla-likos2015}%
  \BibitemOpen
  \bibfield  {author} {\bibinfo {author} {\bibfnamefont {T.}~\bibnamefont
  {Colla}}\ and\ \bibinfo {author} {\bibfnamefont {C.~N.}\ \bibnamefont
  {Likos}},\ }\href@noop {} {\bibfield  {journal} {\bibinfo  {journal} {Mol.
  Phys.}\ }\textbf {\bibinfo {volume} {113}},\ \bibinfo {pages} {2496}
  (\bibinfo {year} {2015})}\BibitemShut {NoStop}%
\bibitem [{\citenamefont {Gupta}\ \emph
  {et~al.}(2015{\natexlab{a}})\citenamefont {Gupta}, \citenamefont {Camargo},
  \citenamefont {Stellbrink}, \citenamefont {Allgaier}, \citenamefont
  {Radulescu}, \citenamefont {Lindner}, \citenamefont {Zaccarelli},
  \citenamefont {Likos},\ and\ \citenamefont
  {Richter}}]{stellbrink-likos-nanoscale2015}%
  \BibitemOpen
  \bibfield  {author} {\bibinfo {author} {\bibfnamefont {S.}~\bibnamefont
  {Gupta}}, \bibinfo {author} {\bibfnamefont {M.}~\bibnamefont {Camargo}},
  \bibinfo {author} {\bibfnamefont {J.}~\bibnamefont {Stellbrink}}, \bibinfo
  {author} {\bibfnamefont {J.}~\bibnamefont {Allgaier}}, \bibinfo {author}
  {\bibfnamefont {A.}~\bibnamefont {Radulescu}}, \bibinfo {author}
  {\bibfnamefont {P.}~\bibnamefont {Lindner}}, \bibinfo {author} {\bibfnamefont
  {E.}~\bibnamefont {Zaccarelli}}, \bibinfo {author} {\bibfnamefont {C.~N.}\
  \bibnamefont {Likos}}, \ and\ \bibinfo {author} {\bibnamefont {Richter}},\
  }\href@noop {} {\bibfield  {journal} {\bibinfo  {journal} {Nanoscale}\
  }\textbf {\bibinfo {volume} {7}},\ \bibinfo {pages} {13924} (\bibinfo {year}
  {2015}{\natexlab{a}})}\BibitemShut {NoStop}%
\bibitem [{\citenamefont {Gupta}\ \emph
  {et~al.}(2015{\natexlab{b}})\citenamefont {Gupta}, \citenamefont
  {Stellbrink}, \citenamefont {Zaccarelli}, \citenamefont {Likos},
  \citenamefont {Camargo}, \citenamefont {Holmqvist}, \citenamefont {Allgaier},
  \citenamefont {Willner},\ and\ \citenamefont
  {Richter}}]{stellbrink-likos-prl2015}%
  \BibitemOpen
  \bibfield  {author} {\bibinfo {author} {\bibfnamefont {S.}~\bibnamefont
  {Gupta}}, \bibinfo {author} {\bibfnamefont {J.}~\bibnamefont {Stellbrink}},
  \bibinfo {author} {\bibfnamefont {E.}~\bibnamefont {Zaccarelli}}, \bibinfo
  {author} {\bibfnamefont {C.~N.}\ \bibnamefont {Likos}}, \bibinfo {author}
  {\bibfnamefont {M.}~\bibnamefont {Camargo}}, \bibinfo {author} {\bibfnamefont
  {P.}~\bibnamefont {Holmqvist}}, \bibinfo {author} {\bibfnamefont
  {J.}~\bibnamefont {Allgaier}}, \bibinfo {author} {\bibfnamefont
  {L.}~\bibnamefont {Willner}}, \ and\ \bibinfo {author} {\bibfnamefont
  {D.}~\bibnamefont {Richter}},\ }\href@noop {} {\bibfield  {journal} {\bibinfo
   {journal} {Phys. Rev. Lett.}\ }\textbf {\bibinfo {volume} {115}},\ \bibinfo
  {pages} {128302} (\bibinfo {year} {2015}{\natexlab{b}})}\BibitemShut
  {NoStop}%
\bibitem [{\citenamefont {Lyon}\ and\ \citenamefont
  {Serpe}(2012)}]{HydrogelBook2012}%
  \BibitemOpen
  \bibinfo {editor} {\bibfnamefont {L.~A.}\ \bibnamefont {Lyon}}\ and\ \bibinfo
  {editor} {\bibfnamefont {M.~J.}\ \bibnamefont {Serpe}},\ eds.,\ \href@noop {}
  {\emph {\bibinfo {title} {Hydrogel Micro and Nanoparticles}}}\ (\bibinfo
  {publisher} {Wiley-VCH Verlag GmbH \& Co. KGaA},\ \bibinfo {address}
  {Weinheim},\ \bibinfo {year} {2012})\BibitemShut {NoStop}%
\bibitem [{\citenamefont {Fern{\'a}ndez-Nieves}\ \emph
  {et~al.}(2011)\citenamefont {Fern{\'a}ndez-Nieves}, \citenamefont {Wyss},
  \citenamefont {Mattsson},\ and\ \citenamefont {Weitz}}]{MicrogelBook2011}%
  \BibitemOpen
  \bibinfo {editor} {\bibfnamefont {A.}~\bibnamefont {Fern{\'a}ndez-Nieves}},
  \bibinfo {editor} {\bibfnamefont {H.}~\bibnamefont {Wyss}}, \bibinfo {editor}
  {\bibfnamefont {J.}~\bibnamefont {Mattsson}}, \ and\ \bibinfo {editor}
  {\bibfnamefont {D.~A.}\ \bibnamefont {Weitz}},\ eds.,\ \href@noop {} {\emph
  {\bibinfo {title} {Microgel Suspensions: Fundamentals and Applications}}}\
  (\bibinfo  {publisher} {Wiley-VCH Verlag GmbH \& Co. KGaA},\ \bibinfo
  {address} {Weinheim},\ \bibinfo {year} {2011})\BibitemShut {NoStop}%
\bibitem [{\citenamefont {Lyon}\ and\ \citenamefont
  {Fern{\'a}ndez-Nieves}(2012)}]{lyon-nieves-AnnuRevPhysChem2012}%
  \BibitemOpen
  \bibfield  {author} {\bibinfo {author} {\bibfnamefont {L.~A.}\ \bibnamefont
  {Lyon}}\ and\ \bibinfo {author} {\bibfnamefont {A.}~\bibnamefont
  {Fern{\'a}ndez-Nieves}},\ }\href@noop {} {\bibfield  {journal} {\bibinfo
  {journal} {Annu. Rev. Phys. Chem.}\ }\textbf {\bibinfo {volume} {63}},\
  \bibinfo {pages} {25} (\bibinfo {year} {2012})}\BibitemShut {NoStop}%
\bibitem [{\citenamefont {Yunker}\ \emph {et~al.}(2014)\citenamefont {Yunker},
  \citenamefont {Chen}, \citenamefont {Gratale}, \citenamefont {Lohr},
  \citenamefont {Still},\ and\ \citenamefont {Yodh}}]{yunker-yodh-review2014}%
  \BibitemOpen
  \bibfield  {author} {\bibinfo {author} {\bibfnamefont {P.~J.}\ \bibnamefont
  {Yunker}}, \bibinfo {author} {\bibfnamefont {K.}~\bibnamefont {Chen}},
  \bibinfo {author} {\bibfnamefont {D.}~\bibnamefont {Gratale}}, \bibinfo
  {author} {\bibfnamefont {M.~A.}\ \bibnamefont {Lohr}}, \bibinfo {author}
  {\bibfnamefont {T.}~\bibnamefont {Still}}, \ and\ \bibinfo {author}
  {\bibfnamefont {A.~G.}\ \bibnamefont {Yodh}},\ }\href@noop {} {\bibfield
  {journal} {\bibinfo  {journal} {Rep. Prog. Phys.}\ }\textbf {\bibinfo
  {volume} {77}},\ \bibinfo {pages} {056601} (\bibinfo {year}
  {2014})}\BibitemShut {NoStop}%
\bibitem [{\citenamefont {Baker}(1949)}]{baker1949}%
  \BibitemOpen
  \bibfield  {author} {\bibinfo {author} {\bibfnamefont {W.~O.}\ \bibnamefont
  {Baker}},\ }\href@noop {} {\bibfield  {journal} {\bibinfo  {journal} {Ind.
  Eng. Chem.}\ }\textbf {\bibinfo {volume} {41}},\ \bibinfo {pages} {511}
  (\bibinfo {year} {1949})}\BibitemShut {NoStop}%
\bibitem [{\citenamefont {Pelton}\ and\ \citenamefont
  {Chibante}(1986)}]{pelton1986}%
  \BibitemOpen
  \bibfield  {author} {\bibinfo {author} {\bibfnamefont {R.~H.}\ \bibnamefont
  {Pelton}}\ and\ \bibinfo {author} {\bibfnamefont {P.}~\bibnamefont
  {Chibante}},\ }\href@noop {} {\bibfield  {journal} {\bibinfo  {journal}
  {Colloids Surf.}\ }\textbf {\bibinfo {volume} {20}},\ \bibinfo {pages} {247}
  (\bibinfo {year} {1986})}\BibitemShut {NoStop}%
\bibitem [{\citenamefont {Pelton}(2000)}]{pelton2000}%
  \BibitemOpen
  \bibfield  {author} {\bibinfo {author} {\bibfnamefont {R.~H.}\ \bibnamefont
  {Pelton}},\ }\href@noop {} {\bibfield  {journal} {\bibinfo  {journal} {Adv.
  Colloid Interface Sci.}\ }\textbf {\bibinfo {volume} {85}},\ \bibinfo {pages}
  {1} (\bibinfo {year} {2000})}\BibitemShut {NoStop}%
\bibitem [{\citenamefont {Saunders}\ \emph {et~al.}(2009)\citenamefont
  {Saunders}, \citenamefont {Laajam}, \citenamefont {Daly}, \citenamefont
  {Teow}, \citenamefont {Hu},\ and\ \citenamefont {Stepto}}]{saunders2009}%
  \BibitemOpen
  \bibfield  {author} {\bibinfo {author} {\bibfnamefont {B.~R.}\ \bibnamefont
  {Saunders}}, \bibinfo {author} {\bibfnamefont {N.}~\bibnamefont {Laajam}},
  \bibinfo {author} {\bibfnamefont {E.}~\bibnamefont {Daly}}, \bibinfo {author}
  {\bibfnamefont {S.}~\bibnamefont {Teow}}, \bibinfo {author} {\bibfnamefont
  {X.}~\bibnamefont {Hu}}, \ and\ \bibinfo {author} {\bibfnamefont
  {R.}~\bibnamefont {Stepto}},\ }\href@noop {} {\bibfield  {journal} {\bibinfo
  {journal} {Adv. Colloid Interface Sci.}\ }\textbf {\bibinfo {volume} {147}},\
  \bibinfo {pages} {251} (\bibinfo {year} {2009})}\BibitemShut {NoStop}%
\bibitem [{\citenamefont {Pelton}\ and\ \citenamefont
  {Hoare}(2011)}]{pelton2011}%
  \BibitemOpen
  \bibfield  {author} {\bibinfo {author} {\bibfnamefont {R.}~\bibnamefont
  {Pelton}}\ and\ \bibinfo {author} {\bibfnamefont {T.}~\bibnamefont {Hoare}},\
  }in\ \href@noop {} {\emph {\bibinfo {booktitle} {Microgel Suspensions:
  Fundamentals and Applications}}},\ \bibinfo {editor} {edited by\ \bibinfo
  {editor} {\bibfnamefont {A.}~\bibnamefont {Fern{\'a}ndez-Nieves}}, \bibinfo
  {editor} {\bibfnamefont {H.}~\bibnamefont {Wyss}}, \bibinfo {editor}
  {\bibfnamefont {J.}~\bibnamefont {Mattsson}}, \ and\ \bibinfo {editor}
  {\bibfnamefont {D.~A.}\ \bibnamefont {Weitz}}}\ (\bibinfo  {publisher}
  {Wiley-VCH Verlag GmbH \& Co. KGaA},\ \bibinfo {address} {Weinheim},\
  \bibinfo {year} {2011})\ pp.\ \bibinfo {pages} {3--32}\BibitemShut {NoStop}%
\bibitem [{\citenamefont {Shah}\ \emph {et~al.}(2008)\citenamefont {Shah},
  \citenamefont {Kim}, \citenamefont {Agresti}, \citenamefont {Weitz},\ and\
  \citenamefont {Chu}}]{weitz-SM2008}%
  \BibitemOpen
  \bibfield  {author} {\bibinfo {author} {\bibfnamefont {R.~K.}\ \bibnamefont
  {Shah}}, \bibinfo {author} {\bibfnamefont {J.-W.}\ \bibnamefont {Kim}},
  \bibinfo {author} {\bibfnamefont {J.~J.}\ \bibnamefont {Agresti}}, \bibinfo
  {author} {\bibfnamefont {D.~A.}\ \bibnamefont {Weitz}}, \ and\ \bibinfo
  {author} {\bibfnamefont {L.-Y.}\ \bibnamefont {Chu}},\ }\href@noop {}
  {\bibfield  {journal} {\bibinfo  {journal} {Soft Matter}\ }\textbf {\bibinfo
  {volume} {4}},\ \bibinfo {pages} {2303} (\bibinfo {year} {2008})}\BibitemShut
  {NoStop}%
\bibitem [{\citenamefont {Still}\ \emph {et~al.}(2013)\citenamefont {Still},
  \citenamefont {Chen}, \citenamefont {Alsayed}, \citenamefont {Aptowicz},\
  and\ \citenamefont {Yodh}}]{yodh2013}%
  \BibitemOpen
  \bibfield  {author} {\bibinfo {author} {\bibfnamefont {T.}~\bibnamefont
  {Still}}, \bibinfo {author} {\bibfnamefont {K.}~\bibnamefont {Chen}},
  \bibinfo {author} {\bibfnamefont {A.~M.}\ \bibnamefont {Alsayed}}, \bibinfo
  {author} {\bibfnamefont {K.~B.}\ \bibnamefont {Aptowicz}}, \ and\ \bibinfo
  {author} {\bibfnamefont {A.~G.}\ \bibnamefont {Yodh}},\ }\href@noop {}
  {\bibfield  {journal} {\bibinfo  {journal} {J. Colloid Interface Sci.}\
  }\textbf {\bibinfo {volume} {405}},\ \bibinfo {pages} {96} (\bibinfo {year}
  {2013})}\BibitemShut {NoStop}%
\bibitem [{\citenamefont {Hamidi}\ \emph {et~al.}(2008)\citenamefont {Hamidi},
  \citenamefont {Azadi},\ and\ \citenamefont {Rafiei}}]{hamidi2008}%
  \BibitemOpen
  \bibfield  {author} {\bibinfo {author} {\bibfnamefont {M.}~\bibnamefont
  {Hamidi}}, \bibinfo {author} {\bibfnamefont {A.}~\bibnamefont {Azadi}}, \
  and\ \bibinfo {author} {\bibfnamefont {P.}~\bibnamefont {Rafiei}},\
  }\href@noop {} {\bibfield  {journal} {\bibinfo  {journal} {Advanced Drug
  Delivery Rev.}\ }\textbf {\bibinfo {volume} {60}},\ \bibinfo {pages} {1638}
  (\bibinfo {year} {2008})}\BibitemShut {NoStop}%
\bibitem [{\citenamefont {Oh}\ \emph {et~al.}(2008)\citenamefont {Oh},
  \citenamefont {Drumright}, \citenamefont {Siegwart},\ and\ \citenamefont
  {Matyjaszewski}}]{oh2008}%
  \BibitemOpen
  \bibfield  {author} {\bibinfo {author} {\bibfnamefont {J.~K.}\ \bibnamefont
  {Oh}}, \bibinfo {author} {\bibfnamefont {R.}~\bibnamefont {Drumright}},
  \bibinfo {author} {\bibfnamefont {D.~J.}\ \bibnamefont {Siegwart}}, \ and\
  \bibinfo {author} {\bibfnamefont {K.}~\bibnamefont {Matyjaszewski}},\
  }\href@noop {} {\bibfield  {journal} {\bibinfo  {journal} {Prog. Polym.
  Sci.}\ }\textbf {\bibinfo {volume} {33}},\ \bibinfo {pages} {448} (\bibinfo
  {year} {2008})}\BibitemShut {NoStop}%
\bibitem [{\citenamefont {Oh}\ \emph {et~al.}(2009)\citenamefont {Oh},
  \citenamefont {Lee},\ and\ \citenamefont {Park}}]{oh2009}%
  \BibitemOpen
  \bibfield  {author} {\bibinfo {author} {\bibfnamefont {J.~K.}\ \bibnamefont
  {Oh}}, \bibinfo {author} {\bibfnamefont {D.~I.}\ \bibnamefont {Lee}}, \ and\
  \bibinfo {author} {\bibfnamefont {J.~M.}\ \bibnamefont {Park}},\ }\href@noop
  {} {\bibfield  {journal} {\bibinfo  {journal} {Prog. Polym. Sci.}\ }\textbf
  {\bibinfo {volume} {34}},\ \bibinfo {pages} {1261} (\bibinfo {year}
  {2009})}\BibitemShut {NoStop}%
\bibitem [{\citenamefont {Schmidt}\ \emph {et~al.}(2011)\citenamefont
  {Schmidt}, \citenamefont {Fernandes}, \citenamefont {Geest}, \citenamefont
  {Delcea}, \citenamefont {Skirtach}, \citenamefont {M{\"o}hwald},\ and\
  \citenamefont {Fery}}]{fery-AdvFunctMater2011}%
  \BibitemOpen
  \bibfield  {author} {\bibinfo {author} {\bibfnamefont {S.}~\bibnamefont
  {Schmidt}}, \bibinfo {author} {\bibfnamefont {P.~A.~L.}\ \bibnamefont
  {Fernandes}}, \bibinfo {author} {\bibfnamefont {B.~G.~D.}\ \bibnamefont
  {Geest}}, \bibinfo {author} {\bibfnamefont {M.}~\bibnamefont {Delcea}},
  \bibinfo {author} {\bibfnamefont {A.~G.}\ \bibnamefont {Skirtach}}, \bibinfo
  {author} {\bibfnamefont {H.}~\bibnamefont {M{\"o}hwald}}, \ and\ \bibinfo
  {author} {\bibfnamefont {A.}~\bibnamefont {Fery}},\ }\href@noop {} {\bibfield
   {journal} {\bibinfo  {journal} {Adv. Funct. Mater.}\ }\textbf {\bibinfo
  {volume} {21}},\ \bibinfo {pages} {1411} (\bibinfo {year}
  {2011})}\BibitemShut {NoStop}%
\bibitem [{\citenamefont {Sivakumaran}\ \emph {et~al.}(2013)\citenamefont
  {Sivakumaran}, \citenamefont {Maitland}, \citenamefont {Oszustowicz},\ and\
  \citenamefont {Hoare}}]{hoare-jcis2013}%
  \BibitemOpen
  \bibfield  {author} {\bibinfo {author} {\bibfnamefont {D.}~\bibnamefont
  {Sivakumaran}}, \bibinfo {author} {\bibfnamefont {D.}~\bibnamefont
  {Maitland}}, \bibinfo {author} {\bibfnamefont {T.}~\bibnamefont
  {Oszustowicz}}, \ and\ \bibinfo {author} {\bibfnamefont {T.}~\bibnamefont
  {Hoare}},\ }\href@noop {} {\bibfield  {journal} {\bibinfo  {journal} {J.
  Colloid Interface Sci.}\ }\textbf {\bibinfo {volume} {392}},\ \bibinfo
  {pages} {422} (\bibinfo {year} {2013})}\BibitemShut {NoStop}%
\bibitem [{\citenamefont {Mohanty}\ and\ \citenamefont
  {Richtering}(2008)}]{mohanty-richtering2008}%
  \BibitemOpen
  \bibfield  {author} {\bibinfo {author} {\bibfnamefont {P.~S.}\ \bibnamefont
  {Mohanty}}\ and\ \bibinfo {author} {\bibfnamefont {W.}~\bibnamefont
  {Richtering}},\ }\href@noop {} {\bibfield  {journal} {\bibinfo  {journal} {J.
  Phys. Chem. B}\ }\textbf {\bibinfo {volume} {112}},\ \bibinfo {pages} {14692}
  (\bibinfo {year} {2008})}\BibitemShut {NoStop}%
\bibitem [{\citenamefont {Eckert}\ and\ \citenamefont
  {Richter}(2008)}]{richtering2008}%
  \BibitemOpen
  \bibfield  {author} {\bibinfo {author} {\bibfnamefont {T.}~\bibnamefont
  {Eckert}}\ and\ \bibinfo {author} {\bibfnamefont {W.}~\bibnamefont
  {Richter}},\ }\href@noop {} {\bibfield  {journal} {\bibinfo  {journal} {J.
  Chem. Phys.}\ }\textbf {\bibinfo {volume} {129}},\ \bibinfo {pages} {124902}
  (\bibinfo {year} {2008})}\BibitemShut {NoStop}%
\bibitem [{\citenamefont {St.~John}\ \emph {et~al.}(2007)\citenamefont
  {St.~John}, \citenamefont {Breedveld},\ and\ \citenamefont
  {Lyon}}]{lyon2007}%
  \BibitemOpen
  \bibfield  {author} {\bibinfo {author} {\bibfnamefont {A.~N.}\ \bibnamefont
  {St.~John}}, \bibinfo {author} {\bibfnamefont {V.}~\bibnamefont {Breedveld}},
  \ and\ \bibinfo {author} {\bibfnamefont {L.~A.}\ \bibnamefont {Lyon}},\
  }\href@noop {} {\bibfield  {journal} {\bibinfo  {journal} {J. Phys. Chem. B}\
  }\textbf {\bibinfo {volume} {111}},\ \bibinfo {pages} {7796} (\bibinfo {year}
  {2007})}\BibitemShut {NoStop}%
\bibitem [{\citenamefont {Muluneh}\ and\ \citenamefont
  {Weitz}(2012)}]{weitz-pre2012}%
  \BibitemOpen
  \bibfield  {author} {\bibinfo {author} {\bibfnamefont {M.}~\bibnamefont
  {Muluneh}}\ and\ \bibinfo {author} {\bibfnamefont {D.~A.}\ \bibnamefont
  {Weitz}},\ }\href@noop {} {\bibfield  {journal} {\bibinfo  {journal} {Phys.
  Rev. E}\ }\textbf {\bibinfo {volume} {85}},\ \bibinfo {pages} {021405}
  (\bibinfo {year} {2012})}\BibitemShut {NoStop}%
\bibitem [{\citenamefont {Riest}\ \emph {et~al.}(2012)\citenamefont {Riest},
  \citenamefont {Mohanty}, \citenamefont {Schurtenberger},\ and\ \citenamefont
  {Likos}}]{schurtenberger-ZPC2012}%
  \BibitemOpen
  \bibfield  {author} {\bibinfo {author} {\bibfnamefont {J.}~\bibnamefont
  {Riest}}, \bibinfo {author} {\bibfnamefont {P.}~\bibnamefont {Mohanty}},
  \bibinfo {author} {\bibfnamefont {P.}~\bibnamefont {Schurtenberger}}, \ and\
  \bibinfo {author} {\bibfnamefont {C.~N.}\ \bibnamefont {Likos}},\ }\href@noop
  {} {\bibfield  {journal} {\bibinfo  {journal} {Z. Phys. Chem.}\ }\textbf
  {\bibinfo {volume} {226}},\ \bibinfo {pages} {711} (\bibinfo {year}
  {2012})}\BibitemShut {NoStop}%
\bibitem [{\citenamefont {Mohanty}\ \emph {et~al.}(2012)\citenamefont
  {Mohanty}, \citenamefont {Yethiraj},\ and\ \citenamefont
  {Schurtenberger}}]{schurtenberger-SM2012}%
  \BibitemOpen
  \bibfield  {author} {\bibinfo {author} {\bibfnamefont {P.~S.}\ \bibnamefont
  {Mohanty}}, \bibinfo {author} {\bibfnamefont {A.}~\bibnamefont {Yethiraj}}, \
  and\ \bibinfo {author} {\bibfnamefont {P.}~\bibnamefont {Schurtenberger}},\
  }\href@noop {} {\bibfield  {journal} {\bibinfo  {journal} {Soft Matter}\
  }\textbf {\bibinfo {volume} {8}},\ \bibinfo {pages} {10819} (\bibinfo {year}
  {2012})}\BibitemShut {NoStop}%
\bibitem [{\citenamefont {Paloli}\ \emph {et~al.}(2013)\citenamefont {Paloli},
  \citenamefont {Mohanty}, \citenamefont {Crassous}, \citenamefont
  {Zaccarelli},\ and\ \citenamefont {Schurtenberger}}]{schurtenberger2013}%
  \BibitemOpen
  \bibfield  {author} {\bibinfo {author} {\bibfnamefont {D.}~\bibnamefont
  {Paloli}}, \bibinfo {author} {\bibfnamefont {P.~S.}\ \bibnamefont {Mohanty}},
  \bibinfo {author} {\bibfnamefont {J.~J.}\ \bibnamefont {Crassous}}, \bibinfo
  {author} {\bibfnamefont {E.}~\bibnamefont {Zaccarelli}}, \ and\ \bibinfo
  {author} {\bibfnamefont {P.}~\bibnamefont {Schurtenberger}},\ }\href@noop {}
  {\bibfield  {journal} {\bibinfo  {journal} {Soft Matter}\ }\textbf {\bibinfo
  {volume} {9}},\ \bibinfo {pages} {3000} (\bibinfo {year} {2013})}\BibitemShut
  {NoStop}%
\bibitem [{\citenamefont {Gasser}\ \emph {et~al.}(2013)\citenamefont {Gasser},
  \citenamefont {Li{\'e}tor-Santos}, \citenamefont {Scotti}, \citenamefont
  {Bunk}, \citenamefont {Menzel},\ and\ \citenamefont
  {Fern{\'a}ndez-Nieves}}]{nieves-pre2013}%
  \BibitemOpen
  \bibfield  {author} {\bibinfo {author} {\bibfnamefont {U.}~\bibnamefont
  {Gasser}}, \bibinfo {author} {\bibfnamefont {J.-J.}\ \bibnamefont
  {Li{\'e}tor-Santos}}, \bibinfo {author} {\bibfnamefont {A.}~\bibnamefont
  {Scotti}}, \bibinfo {author} {\bibfnamefont {O.}~\bibnamefont {Bunk}},
  \bibinfo {author} {\bibfnamefont {A.}~\bibnamefont {Menzel}}, \ and\ \bibinfo
  {author} {\bibfnamefont {A.}~\bibnamefont {Fern{\'a}ndez-Nieves}},\
  }\href@noop {} {\bibfield  {journal} {\bibinfo  {journal} {Phys. Rev. E}\
  }\textbf {\bibinfo {volume} {88}},\ \bibinfo {pages} {052308} (\bibinfo
  {year} {2013})}\BibitemShut {NoStop}%
\bibitem [{\citenamefont {Mohanty}\ \emph {et~al.}(2014)\citenamefont
  {Mohanty}, \citenamefont {Paloli}, \citenamefont {Crassous}, \citenamefont
  {Zaccarelli},\ and\ \citenamefont {Schurtenberger}}]{schurtenberger2014}%
  \BibitemOpen
  \bibfield  {author} {\bibinfo {author} {\bibfnamefont {P.~S.}\ \bibnamefont
  {Mohanty}}, \bibinfo {author} {\bibfnamefont {D.}~\bibnamefont {Paloli}},
  \bibinfo {author} {\bibfnamefont {J.~J.}\ \bibnamefont {Crassous}}, \bibinfo
  {author} {\bibfnamefont {E.}~\bibnamefont {Zaccarelli}}, \ and\ \bibinfo
  {author} {\bibfnamefont {P.}~\bibnamefont {Schurtenberger}},\ }\href@noop {}
  {\bibfield  {journal} {\bibinfo  {journal} {J. Chem. Phys.}\ }\textbf
  {\bibinfo {volume} {140}},\ \bibinfo {pages} {094901} (\bibinfo {year}
  {2014})}\BibitemShut {NoStop}%
\bibitem [{\citenamefont {Denton}(2003)}]{denton2003}%
  \BibitemOpen
  \bibfield  {author} {\bibinfo {author} {\bibfnamefont {A.~R.}\ \bibnamefont
  {Denton}},\ }\href@noop {} {\bibfield  {journal} {\bibinfo  {journal} {Phys.
  Rev. E}\ }\textbf {\bibinfo {volume} {67}},\ \bibinfo {pages} {011804}
  (\bibinfo {year} {2003})},\ \bibinfo {note} {{\bf 68}, 049904(E)
  (2003)}\BibitemShut {NoStop}%
\bibitem [{\citenamefont {Gottwald}\ \emph {et~al.}(2005)\citenamefont
  {Gottwald}, \citenamefont {Likos}, \citenamefont {Kahl},\ and\ \citenamefont
  {L{\"o}wen}}]{gottwald2005}%
  \BibitemOpen
  \bibfield  {author} {\bibinfo {author} {\bibfnamefont {D.}~\bibnamefont
  {Gottwald}}, \bibinfo {author} {\bibfnamefont {C.~N.}\ \bibnamefont {Likos}},
  \bibinfo {author} {\bibfnamefont {G.}~\bibnamefont {Kahl}}, \ and\ \bibinfo
  {author} {\bibfnamefont {H.}~\bibnamefont {L{\"o}wen}},\ }\href@noop {}
  {\bibfield  {journal} {\bibinfo  {journal} {J. Chem. Phys.}\ }\textbf
  {\bibinfo {volume} {122}},\ \bibinfo {pages} {074903} (\bibinfo {year}
  {2005})}\BibitemShut {NoStop}%
\bibitem [{\citenamefont {Likos}(2011)}]{likos2011}%
  \BibitemOpen
  \bibfield  {author} {\bibinfo {author} {\bibfnamefont {C.~N.}\ \bibnamefont
  {Likos}},\ }in\ \href@noop {} {\emph {\bibinfo {booktitle} {Microgel
  Suspensions: Fundamentals and Applications}}},\ \bibinfo {editor} {edited by\
  \bibinfo {editor} {\bibfnamefont {A.}~\bibnamefont {Fern{\'a}ndez-Nieves}},
  \bibinfo {editor} {\bibfnamefont {H.}~\bibnamefont {Wyss}}, \bibinfo {editor}
  {\bibfnamefont {J.}~\bibnamefont {Mattsson}}, \ and\ \bibinfo {editor}
  {\bibfnamefont {D.~A.}\ \bibnamefont {Weitz}}}\ (\bibinfo  {publisher}
  {Wiley-VCH Verlag GmbH \& Co. KGaA},\ \bibinfo {address} {Weinheim},\
  \bibinfo {year} {2011})\ pp.\ \bibinfo {pages} {165--193}\BibitemShut
  {NoStop}%
\bibitem [{\citenamefont {Baulin}\ and\ \citenamefont
  {Trizac}(2012)}]{trizac2012}%
  \BibitemOpen
  \bibfield  {author} {\bibinfo {author} {\bibfnamefont {V.~A.}\ \bibnamefont
  {Baulin}}\ and\ \bibinfo {author} {\bibfnamefont {E.}~\bibnamefont
  {Trizac}},\ }\href@noop {} {\bibfield  {journal} {\bibinfo  {journal} {Soft
  Matter}\ }\textbf {\bibinfo {volume} {8}},\ \bibinfo {pages} {6755} (\bibinfo
  {year} {2012})}\BibitemShut {NoStop}%
\bibitem [{\citenamefont {Hedrick}\ \emph {et~al.}(2015)\citenamefont
  {Hedrick}, \citenamefont {Chung},\ and\ \citenamefont
  {Denton}}]{hedrick-chung-denton2015}%
  \BibitemOpen
  \bibfield  {author} {\bibinfo {author} {\bibfnamefont {M.~M.}\ \bibnamefont
  {Hedrick}}, \bibinfo {author} {\bibfnamefont {J.~K.}\ \bibnamefont {Chung}},
  \ and\ \bibinfo {author} {\bibfnamefont {A.~R.}\ \bibnamefont {Denton}},\
  }\href@noop {} {\bibfield  {journal} {\bibinfo  {journal} {J. Chem. Phys.}\
  }\textbf {\bibinfo {volume} {142}},\ \bibinfo {pages} {034904} (\bibinfo
  {year} {2015})}\BibitemShut {NoStop}%
\bibitem [{\citenamefont {Claudio}\ \emph {et~al.}(2009)\citenamefont
  {Claudio}, \citenamefont {Kremer},\ and\ \citenamefont {Holm}}]{holm2009}%
  \BibitemOpen
  \bibfield  {author} {\bibinfo {author} {\bibfnamefont {G.~C.}\ \bibnamefont
  {Claudio}}, \bibinfo {author} {\bibfnamefont {K.}~\bibnamefont {Kremer}}, \
  and\ \bibinfo {author} {\bibfnamefont {C.}~\bibnamefont {Holm}},\ }\href@noop
  {} {\bibfield  {journal} {\bibinfo  {journal} {J. Chem. Phys.}\ }\textbf
  {\bibinfo {volume} {131}},\ \bibinfo {pages} {094903} (\bibinfo {year}
  {2009})}\BibitemShut {NoStop}%
\bibitem [{\citenamefont {Quesada-P{\'e}rez}\ and\ \citenamefont
  {Mart{\'i}n-Molina}(2013)}]{molina2013}%
  \BibitemOpen
  \bibfield  {author} {\bibinfo {author} {\bibfnamefont {M.}~\bibnamefont
  {Quesada-P{\'e}rez}}\ and\ \bibinfo {author} {\bibfnamefont {A.}~\bibnamefont
  {Mart{\'i}n-Molina}},\ }\href@noop {} {\bibfield  {journal} {\bibinfo
  {journal} {Soft Matter}\ }\textbf {\bibinfo {volume} {9}},\ \bibinfo {pages}
  {7086} (\bibinfo {year} {2013})}\BibitemShut {NoStop}%
\bibitem [{\citenamefont {Schneider}\ and\ \citenamefont
  {Linse}(2002)}]{linse2002}%
  \BibitemOpen
  \bibfield  {author} {\bibinfo {author} {\bibfnamefont {S.}~\bibnamefont
  {Schneider}}\ and\ \bibinfo {author} {\bibfnamefont {P.}~\bibnamefont
  {Linse}},\ }\href@noop {} {\bibfield  {journal} {\bibinfo  {journal} {Eur.
  Phys. J. E}\ }\textbf {\bibinfo {volume} {8}},\ \bibinfo {pages} {457}
  (\bibinfo {year} {2002})}\BibitemShut {NoStop}%
\bibitem [{\citenamefont {Kobayashi}\ and\ \citenamefont
  {Winkler}(2014)}]{winkler2014}%
  \BibitemOpen
  \bibfield  {author} {\bibinfo {author} {\bibfnamefont {H.}~\bibnamefont
  {Kobayashi}}\ and\ \bibinfo {author} {\bibfnamefont {R.~G.}\ \bibnamefont
  {Winkler}},\ }\href@noop {} {\bibfield  {journal} {\bibinfo  {journal}
  {Polymers}\ }\textbf {\bibinfo {volume} {6}},\ \bibinfo {pages} {1602}
  (\bibinfo {year} {2014})}\BibitemShut {NoStop}%
\bibitem [{\citenamefont {Kobayashi}\ and\ \citenamefont
  {Winkler}(2016)}]{winkler2016}%
  \BibitemOpen
  \bibfield  {author} {\bibinfo {author} {\bibfnamefont {H.}~\bibnamefont
  {Kobayashi}}\ and\ \bibinfo {author} {\bibfnamefont {R.~G.}\ \bibnamefont
  {Winkler}},\ }\href@noop {} {\bibfield  {journal} {\bibinfo  {journal} {Sci.
  Rep.}\ }\textbf {\bibinfo {volume} {6}},\ \bibinfo {pages} {19836} (\bibinfo
  {year} {2016})}\BibitemShut {NoStop}%
\bibitem [{\citenamefont {Hooper}\ \emph {et~al.}(1990)\citenamefont {Hooper},
  \citenamefont {Baker}, \citenamefont {Blanch},\ and\ \citenamefont
  {Prausnitz}}]{prausnitz1990}%
  \BibitemOpen
  \bibfield  {author} {\bibinfo {author} {\bibfnamefont {H.~H.}\ \bibnamefont
  {Hooper}}, \bibinfo {author} {\bibfnamefont {J.~P.}\ \bibnamefont {Baker}},
  \bibinfo {author} {\bibfnamefont {H.~W.}\ \bibnamefont {Blanch}}, \ and\
  \bibinfo {author} {\bibfnamefont {J.~M.}\ \bibnamefont {Prausnitz}},\
  }\href@noop {} {\bibfield  {journal} {\bibinfo  {journal} {Macromol.}\
  }\textbf {\bibinfo {volume} {23}},\ \bibinfo {pages} {1096} (\bibinfo {year}
  {1990})}\BibitemShut {NoStop}%
\bibitem [{\citenamefont {English}\ \emph {et~al.}(1996)\citenamefont
  {English}, \citenamefont {Maf{\'e}}, \citenamefont {Manzanares},
  \citenamefont {Yu}, \citenamefont {Grosberg},\ and\ \citenamefont
  {Tanaka}}]{english-grosberg1996}%
  \BibitemOpen
  \bibfield  {author} {\bibinfo {author} {\bibfnamefont {A.~E.}\ \bibnamefont
  {English}}, \bibinfo {author} {\bibfnamefont {S.}~\bibnamefont {Maf{\'e}}},
  \bibinfo {author} {\bibfnamefont {J.~A.}\ \bibnamefont {Manzanares}},
  \bibinfo {author} {\bibfnamefont {X.}~\bibnamefont {Yu}}, \bibinfo {author}
  {\bibfnamefont {A.~Y.}\ \bibnamefont {Grosberg}}, \ and\ \bibinfo {author}
  {\bibfnamefont {T.}~\bibnamefont {Tanaka}},\ }\href@noop {} {\bibfield
  {journal} {\bibinfo  {journal} {J. Chem. Phys.}\ }\textbf {\bibinfo {volume}
  {104}},\ \bibinfo {pages} {8713} (\bibinfo {year} {1996})}\BibitemShut
  {NoStop}%
\bibitem [{\citenamefont {Nisato}\ \emph {et~al.}(1999)\citenamefont {Nisato},
  \citenamefont {Munch},\ and\ \citenamefont {Candau}}]{nisato-langmuir1999}%
  \BibitemOpen
  \bibfield  {author} {\bibinfo {author} {\bibfnamefont {G.}~\bibnamefont
  {Nisato}}, \bibinfo {author} {\bibfnamefont {J.~P.}\ \bibnamefont {Munch}}, \
  and\ \bibinfo {author} {\bibfnamefont {S.~J.}\ \bibnamefont {Candau}},\
  }\href@noop {} {\bibfield  {journal} {\bibinfo  {journal} {Langmuir}\
  }\textbf {\bibinfo {volume} {15}},\ \bibinfo {pages} {4236} (\bibinfo {year}
  {1999})}\BibitemShut {NoStop}%
\bibitem [{\citenamefont {Fern{\'a}ndez-Nieves}\ \emph
  {et~al.}(2001)\citenamefont {Fern{\'a}ndez-Nieves}, \citenamefont
  {Fern{\'a}ndez-Barbero},\ and\ \citenamefont {de~las
  {Nieves}}}]{nieves-jcp2001}%
  \BibitemOpen
  \bibfield  {author} {\bibinfo {author} {\bibfnamefont {A.}~\bibnamefont
  {Fern{\'a}ndez-Nieves}}, \bibinfo {author} {\bibfnamefont {A.}~\bibnamefont
  {Fern{\'a}ndez-Barbero}}, \ and\ \bibinfo {author} {\bibfnamefont {F.~J.}\
  \bibnamefont {de~las {Nieves}}},\ }\href@noop {} {\bibfield  {journal}
  {\bibinfo  {journal} {J. Chem. Phys.}\ }\textbf {\bibinfo {volume} {115}},\
  \bibinfo {pages} {7644} (\bibinfo {year} {2001})}\BibitemShut {NoStop}%
\bibitem [{\citenamefont {L{\'o}pez-Le{\'o}n}\ \emph
  {et~al.}(2006)\citenamefont {L{\'o}pez-Le{\'o}n}, \citenamefont
  {Ortega-Vinuesa}, \citenamefont {Bastos-Gonz{\'a}lez},\ and\ \citenamefont
  {Ela{\"i}ssari}}]{ortega-vinuesa2006}%
  \BibitemOpen
  \bibfield  {author} {\bibinfo {author} {\bibfnamefont {T.}~\bibnamefont
  {L{\'o}pez-Le{\'o}n}}, \bibinfo {author} {\bibfnamefont {J.~L.}\ \bibnamefont
  {Ortega-Vinuesa}}, \bibinfo {author} {\bibfnamefont {D.}~\bibnamefont
  {Bastos-Gonz{\'a}lez}}, \ and\ \bibinfo {author} {\bibfnamefont
  {A.}~\bibnamefont {Ela{\"i}ssari}},\ }\href@noop {} {\bibfield  {journal}
  {\bibinfo  {journal} {J. Phys. Chem. B}\ }\textbf {\bibinfo {volume} {110}},\
  \bibinfo {pages} {4629} (\bibinfo {year} {2006})}\BibitemShut {NoStop}%
\bibitem [{\citenamefont {Capriles-Gonz{\'a}lez}\ \emph
  {et~al.}(2008)\citenamefont {Capriles-Gonz{\'a}lez}, \citenamefont
  {Sierra-Mart{\'i}n}, \citenamefont {Fern{\'a}ndez-Nieves},\ and\
  \citenamefont {Fern{\'a}ndez-Barbero}}]{nieves-salt-jpcb2008}%
  \BibitemOpen
  \bibfield  {author} {\bibinfo {author} {\bibfnamefont {D.}~\bibnamefont
  {Capriles-Gonz{\'a}lez}}, \bibinfo {author} {\bibfnamefont {B.}~\bibnamefont
  {Sierra-Mart{\'i}n}}, \bibinfo {author} {\bibfnamefont {A.}~\bibnamefont
  {Fern{\'a}ndez-Nieves}}, \ and\ \bibinfo {author} {\bibfnamefont
  {A.}~\bibnamefont {Fern{\'a}ndez-Barbero}},\ }\href@noop {} {\bibfield
  {journal} {\bibinfo  {journal} {J. Phys. Chem. B}\ }\textbf {\bibinfo
  {volume} {112}},\ \bibinfo {pages} {12195} (\bibinfo {year}
  {2008})}\BibitemShut {NoStop}%
\bibitem [{\citenamefont {Flory}(1953)}]{flory1953}%
  \BibitemOpen
  \bibfield  {author} {\bibinfo {author} {\bibfnamefont {P.~J.}\ \bibnamefont
  {Flory}},\ }\href@noop {} {\emph {\bibinfo {title} {Principles of Polymer
  Chemistry}}}\ (\bibinfo  {publisher} {Cornell University Press},\ \bibinfo
  {address} {Ithaca},\ \bibinfo {year} {1953})\BibitemShut {NoStop}%
\bibitem [{\citenamefont {Katchalsky}\ \emph {et~al.}(1951)\citenamefont
  {Katchalsky}, \citenamefont {Lifson},\ and\ \citenamefont
  {Eisenberg}}]{katchalsky1951}%
  \BibitemOpen
  \bibfield  {author} {\bibinfo {author} {\bibfnamefont {A.}~\bibnamefont
  {Katchalsky}}, \bibinfo {author} {\bibfnamefont {S.}~\bibnamefont {Lifson}},
  \ and\ \bibinfo {author} {\bibfnamefont {H.}~\bibnamefont {Eisenberg}},\
  }\href@noop {} {\bibfield  {journal} {\bibinfo  {journal} {J. Polym. Sci. A}\
  }\textbf {\bibinfo {volume} {7}},\ \bibinfo {pages} {571} (\bibinfo {year}
  {1951})}\BibitemShut {NoStop}%
\bibitem [{\citenamefont {Katchalsky}\ and\ \citenamefont
  {Michaeli}(1955)}]{katchalsky1955}%
  \BibitemOpen
  \bibfield  {author} {\bibinfo {author} {\bibfnamefont {A.}~\bibnamefont
  {Katchalsky}}\ and\ \bibinfo {author} {\bibfnamefont {I.}~\bibnamefont
  {Michaeli}},\ }\href@noop {} {\bibfield  {journal} {\bibinfo  {journal} {J.
  Polym. Sci. A}\ }\textbf {\bibinfo {volume} {15}},\ \bibinfo {pages} {69}
  (\bibinfo {year} {1955})}\BibitemShut {NoStop}%
\bibitem [{\citenamefont {de~Gennes}(1979)}]{deGennes1979}%
  \BibitemOpen
  \bibfield  {author} {\bibinfo {author} {\bibfnamefont {P.-G.}\ \bibnamefont
  {de~Gennes}},\ }\href@noop {} {\emph {\bibinfo {title} {Scaling Concepts in
  Polymer Physics}}}\ (\bibinfo  {publisher} {Cornell},\ \bibinfo {address}
  {Ithaca},\ \bibinfo {year} {1979})\BibitemShut {NoStop}%
\bibitem [{\citenamefont {Barrat}\ \emph {et~al.}(1992)\citenamefont {Barrat},
  \citenamefont {Joanny},\ and\ \citenamefont
  {Pincus}}]{barrat-joanny-pincus1992}%
  \BibitemOpen
  \bibfield  {author} {\bibinfo {author} {\bibfnamefont {J.-L.}\ \bibnamefont
  {Barrat}}, \bibinfo {author} {\bibfnamefont {J.-F.}\ \bibnamefont {Joanny}},
  \ and\ \bibinfo {author} {\bibfnamefont {P.}~\bibnamefont {Pincus}},\
  }\href@noop {} {\bibfield  {journal} {\bibinfo  {journal} {J. Phys. {II}
  France}\ }\textbf {\bibinfo {volume} {2}},\ \bibinfo {pages} {1531} (\bibinfo
  {year} {1992})}\BibitemShut {NoStop}%
\bibitem [{\citenamefont {Rubinstein}\ \emph {et~al.}(1996)\citenamefont
  {Rubinstein}, \citenamefont {Colby}, \citenamefont {Dobrynin},\ and\
  \citenamefont {Joanny}}]{rubinstein-dobrynin1996}%
  \BibitemOpen
  \bibfield  {author} {\bibinfo {author} {\bibfnamefont {M.}~\bibnamefont
  {Rubinstein}}, \bibinfo {author} {\bibfnamefont {R.~H.}\ \bibnamefont
  {Colby}}, \bibinfo {author} {\bibfnamefont {A.~V.}\ \bibnamefont {Dobrynin}},
  \ and\ \bibinfo {author} {\bibfnamefont {J.-F.}\ \bibnamefont {Joanny}},\
  }\href@noop {} {\bibfield  {journal} {\bibinfo  {journal} {Macromol.}\
  }\textbf {\bibinfo {volume} {29}},\ \bibinfo {pages} {398} (\bibinfo {year}
  {1996})}\BibitemShut {NoStop}%
\bibitem [{\citenamefont {Kramarenko}\ \emph {et~al.}(1997)\citenamefont
  {Kramarenko}, \citenamefont {Khokhlov},\ and\ \citenamefont
  {Yoshikawa}}]{khokhlov1997}%
  \BibitemOpen
  \bibfield  {author} {\bibinfo {author} {\bibfnamefont {E.~Y.}\ \bibnamefont
  {Kramarenko}}, \bibinfo {author} {\bibfnamefont {A.~R.}\ \bibnamefont
  {Khokhlov}}, \ and\ \bibinfo {author} {\bibfnamefont {K.}~\bibnamefont
  {Yoshikawa}},\ }\href@noop {} {\bibfield  {journal} {\bibinfo  {journal}
  {Macromol.}\ }\textbf {\bibinfo {volume} {30}},\ \bibinfo {pages} {3383}
  (\bibinfo {year} {1997})}\BibitemShut {NoStop}%
\bibitem [{\citenamefont {Malmsten}(2011)}]{malmsten2011}%
  \BibitemOpen
  \bibfield  {author} {\bibinfo {author} {\bibfnamefont {M.}~\bibnamefont
  {Malmsten}},\ }in\ \href@noop {} {\emph {\bibinfo {booktitle} {Microgel
  Suspensions: Fundamentals and Applications}}},\ \bibinfo {editor} {edited by\
  \bibinfo {editor} {\bibfnamefont {A.}~\bibnamefont {Fern{\'a}ndez-Nieves}},
  \bibinfo {editor} {\bibfnamefont {H.}~\bibnamefont {Wyss}}, \bibinfo {editor}
  {\bibfnamefont {J.}~\bibnamefont {Mattsson}}, \ and\ \bibinfo {editor}
  {\bibfnamefont {D.~A.}\ \bibnamefont {Weitz}}}\ (\bibinfo  {publisher}
  {Wiley-VCH Verlag GmbH \& Co. KGaA},\ \bibinfo {address} {Weinheim},\
  \bibinfo {year} {2011})\ pp.\ \bibinfo {pages} {375--405}\BibitemShut
  {NoStop}%
\bibitem [{\citenamefont {Nordstrom}\ \emph {et~al.}(2010)\citenamefont
  {Nordstrom}, \citenamefont {Verneuil}, \citenamefont {Arratia}, \citenamefont
  {Basu}, \citenamefont {Zhang}, \citenamefont {Yodh}, \citenamefont {Gollub},\
  and\ \citenamefont {Durian}}]{nordstrom2010}%
  \BibitemOpen
  \bibfield  {author} {\bibinfo {author} {\bibfnamefont {K.~N.}\ \bibnamefont
  {Nordstrom}}, \bibinfo {author} {\bibfnamefont {E.}~\bibnamefont {Verneuil}},
  \bibinfo {author} {\bibfnamefont {P.~E.}\ \bibnamefont {Arratia}}, \bibinfo
  {author} {\bibfnamefont {A.}~\bibnamefont {Basu}}, \bibinfo {author}
  {\bibfnamefont {Z.}~\bibnamefont {Zhang}}, \bibinfo {author} {\bibfnamefont
  {A.~G.}\ \bibnamefont {Yodh}}, \bibinfo {author} {\bibfnamefont {J.~P.}\
  \bibnamefont {Gollub}}, \ and\ \bibinfo {author} {\bibfnamefont {D.~J.}\
  \bibnamefont {Durian}},\ }\href@noop {} {\bibfield  {journal} {\bibinfo
  {journal} {\PRL}\ }\textbf {\bibinfo {volume} {105}},\ \bibinfo {pages}
  {175701} (\bibinfo {year} {2010})}\BibitemShut {NoStop}%
\bibitem [{\citenamefont {Bouchoux}\ \emph {et~al.}(2013)\citenamefont
  {Bouchoux}, \citenamefont {Qu}, \citenamefont {Bacchin},\ and\ \citenamefont
  {G{\'e}san-Guiziou}}]{bacchin2014}%
  \BibitemOpen
  \bibfield  {author} {\bibinfo {author} {\bibfnamefont {A.}~\bibnamefont
  {Bouchoux}}, \bibinfo {author} {\bibfnamefont {P.}~\bibnamefont {Qu}},
  \bibinfo {author} {\bibfnamefont {P.}~\bibnamefont {Bacchin}}, \ and\
  \bibinfo {author} {\bibfnamefont {G.}~\bibnamefont {G{\'e}san-Guiziou}},\
  }\href@noop {} {\bibfield  {journal} {\bibinfo  {journal} {Langmuir}\
  }\textbf {\bibinfo {volume} {30}},\ \bibinfo {pages} {22} (\bibinfo {year}
  {2013})}\BibitemShut {NoStop}%
\bibitem [{\citenamefont {Roa}\ \emph {et~al.}(2015)\citenamefont {Roa},
  \citenamefont {Zholkovskiy},\ and\ \citenamefont
  {N{\"a}gele}}]{roa-naegele2015}%
  \BibitemOpen
  \bibfield  {author} {\bibinfo {author} {\bibfnamefont {R.}~\bibnamefont
  {Roa}}, \bibinfo {author} {\bibfnamefont {E.~K.}\ \bibnamefont
  {Zholkovskiy}}, \ and\ \bibinfo {author} {\bibfnamefont {G.}~\bibnamefont
  {N{\"a}gele}},\ }\href@noop {} {\bibfield  {journal} {\bibinfo  {journal}
  {Soft Matter}\ }\textbf {\bibinfo {volume} {11}},\ \bibinfo {pages} {4106}
  (\bibinfo {year} {2015})}\BibitemShut {NoStop}%
\bibitem [{\citenamefont {Manning}(1969)}]{manning1969}%
  \BibitemOpen
  \bibfield  {author} {\bibinfo {author} {\bibfnamefont {G.~S.}\ \bibnamefont
  {Manning}},\ }\href@noop {} {\bibfield  {journal} {\bibinfo  {journal} {J.
  Chem. Phys.}\ }\textbf {\bibinfo {volume} {51}},\ \bibinfo {pages} {924}
  (\bibinfo {year} {1969})}\BibitemShut {NoStop}%
\bibitem [{\citenamefont {Marcus}(1955)}]{marcus1955}%
  \BibitemOpen
  \bibfield  {author} {\bibinfo {author} {\bibfnamefont {R.~A.}\ \bibnamefont
  {Marcus}},\ }\href@noop {} {\bibfield  {journal} {\bibinfo  {journal} {J.
  Chem. Phys.}\ }\textbf {\bibinfo {volume} {23}},\ \bibinfo {pages} {1057}
  (\bibinfo {year} {1955})}\BibitemShut {NoStop}%
\bibitem [{\citenamefont {Denton}(2010)}]{denton2010}%
  \BibitemOpen
  \bibfield  {author} {\bibinfo {author} {\bibfnamefont {A.~R.}\ \bibnamefont
  {Denton}},\ }\href@noop {} {\bibfield  {journal} {\bibinfo  {journal} {J.
  Phys.: Condens. Matter}\ }\textbf {\bibinfo {volume} {22}},\ \bibinfo {pages}
  {364108} (\bibinfo {year} {2010})}\BibitemShut {NoStop}%
\bibitem [{\citenamefont {Hallez}\ \emph {et~al.}(2014)\citenamefont {Hallez},
  \citenamefont {Diatta},\ and\ \citenamefont {Meireles}}]{hallez2014}%
  \BibitemOpen
  \bibfield  {author} {\bibinfo {author} {\bibfnamefont {Y.}~\bibnamefont
  {Hallez}}, \bibinfo {author} {\bibfnamefont {J.}~\bibnamefont {Diatta}}, \
  and\ \bibinfo {author} {\bibfnamefont {M.}~\bibnamefont {Meireles}},\
  }\href@noop {} {\bibfield  {journal} {\bibinfo  {journal} {Langmuir}\
  }\textbf {\bibinfo {volume} {30}},\ \bibinfo {pages} {6721} (\bibinfo {year}
  {2014})}\BibitemShut {NoStop}%
\bibitem [{\citenamefont {Urich}\ and\ \citenamefont
  {Denton}()}]{urich-denton2016}%
  \BibitemOpen
  \bibfield  {author} {\bibinfo {author} {\bibfnamefont {M.}~\bibnamefont
  {Urich}}\ and\ \bibinfo {author} {\bibfnamefont {A.~R.}\ \bibnamefont
  {Denton}},\ }\href@noop {} {\bibinfo  {journal} {unpublished}\ }\BibitemShut
  {NoStop}%
\bibitem [{\citenamefont {Wennerstr{\"o}m}\ \emph {et~al.}(1982)\citenamefont
  {Wennerstr{\"o}m}, \citenamefont {J{\"o}nsson},\ and\ \citenamefont
  {Linse}}]{wennerstrom1982}%
  \BibitemOpen
\bibfield  {journal} {  }\bibfield  {author} {\bibinfo {author} {\bibfnamefont
  {H.}~\bibnamefont {Wennerstr{\"o}m}}, \bibinfo {author} {\bibfnamefont
  {B.}~\bibnamefont {J{\"o}nsson}}, \ and\ \bibinfo {author} {\bibfnamefont
  {P.}~\bibnamefont {Linse}},\ }\href@noop {} {\bibfield  {journal} {\bibinfo
  {journal} {J. Chem. Phys.}\ }\textbf {\bibinfo {volume} {76}},\ \bibinfo
  {pages} {4665} (\bibinfo {year} {1982})}\BibitemShut {NoStop}%
\bibitem [{\citenamefont {Deserno}\ and\ \citenamefont
  {Holm}(2001)}]{deserno-holm2001}%
  \BibitemOpen
  \bibfield  {author} {\bibinfo {author} {\bibfnamefont {M.}~\bibnamefont
  {Deserno}}\ and\ \bibinfo {author} {\bibfnamefont {C.}~\bibnamefont {Holm}},\
  }in\ \href@noop {} {\emph {\bibinfo {booktitle} {Electrostatic Effects in
  Soft Matter and Biophysics}}},\ \bibinfo {series} {NATO Advanced Studies
  Institute, Series {II}: {M}athematics Physics and Chemistry}, Vol.~\bibinfo
  {volume} {46},\ \bibinfo {editor} {edited by\ \bibinfo {editor}
  {\bibfnamefont {C.}~\bibnamefont {Holm}}, \bibinfo {editor} {\bibfnamefont
  {P.}~\bibnamefont {K{\'e}kicheff}}, \ and\ \bibinfo {editor} {\bibfnamefont
  {R.}~\bibnamefont {Podgornik}}}\ (\bibinfo  {publisher} {Kluwer},\ \bibinfo
  {address} {Dordrecht},\ \bibinfo {year} {2001})\ pp.\ \bibinfo {pages}
  {27--50}\BibitemShut {NoStop}%
\bibitem [{\citenamefont {Plimpton}(1995)}]{plimpton1995}%
  \BibitemOpen
  \bibfield  {author} {\bibinfo {author} {\bibfnamefont {S.}~\bibnamefont
  {Plimpton}},\ }\href@noop {} {\bibfield  {journal} {\bibinfo  {journal} {J.
  Comp. Phys.}\ }\textbf {\bibinfo {volume} {117}},\ \bibinfo {pages} {1}
  (\bibinfo {year} {1995})}\BibitemShut {NoStop}%
\bibitem [{lam()}]{lammps}%
  \BibitemOpen
  \href@noop {} {}\bibinfo {howpublished}
  {\url{http://lammps.sandia.gov}}\BibitemShut {NoStop}%
\bibitem [{\citenamefont {Stieger}\ \emph {et~al.}(2004)\citenamefont
  {Stieger}, \citenamefont {Richtering}, \citenamefont {Pedersen},\ and\
  \citenamefont {Lindner}}]{stieger2004}%
  \BibitemOpen
  \bibfield  {author} {\bibinfo {author} {\bibfnamefont {M.}~\bibnamefont
  {Stieger}}, \bibinfo {author} {\bibfnamefont {W.}~\bibnamefont {Richtering}},
  \bibinfo {author} {\bibfnamefont {J.~S.}\ \bibnamefont {Pedersen}}, \ and\
  \bibinfo {author} {\bibfnamefont {P.}~\bibnamefont {Lindner}},\ }\href@noop
  {} {\bibfield  {journal} {\bibinfo  {journal} {J. Chem. Phys.}\ }\textbf
  {\bibinfo {volume} {120}},\ \bibinfo {pages} {6197} (\bibinfo {year}
  {2004})}\BibitemShut {NoStop}%
\bibitem [{\citenamefont {Rumyantsev}\ \emph {et~al.}(2015)\citenamefont
  {Rumyantsev}, \citenamefont {Rudov},\ and\ \citenamefont
  {Potemkin}}]{potemkin2015}%
  \BibitemOpen
  \bibfield  {author} {\bibinfo {author} {\bibfnamefont {A.~M.}\ \bibnamefont
  {Rumyantsev}}, \bibinfo {author} {\bibfnamefont {A.~A.}\ \bibnamefont
  {Rudov}}, \ and\ \bibinfo {author} {\bibfnamefont {I.~I.}\ \bibnamefont
  {Potemkin}},\ }\href@noop {} {\bibfield  {journal} {\bibinfo  {journal} {J.
  Chem. Phys.}\ }\textbf {\bibinfo {volume} {142}},\ \bibinfo {pages} {171105}
  (\bibinfo {year} {2015})}\BibitemShut {NoStop}%
\end{thebibliography}

%

\end{document}